\pdfoutput=1
\documentclass[twocolumn]{emulateapj}
\usepackage{graphicx}
\usepackage{amsmath}
\usepackage{appendix} 
\usepackage{float}
\usepackage{placeins}

\begin{document}

\newcommand{\effall}{7\%}
\newcommand{\effrem}{6\%}
\newcommand{\effremtwo}{11\%}

\newcommand{\remnant}{0509$-$67.5}
\newcommand{\remtwo}{0519$-$69.0}
\newcommand{\fsv}{$6500\pm200$ km s$^{-1}$}
\newcommand{\stat}{$\chi^2$}
\newcommand{\ha}{H$\alpha$}
\newcommand{\HST}{{\em HST}}
\newcommand{\betaone}{$0.04$}
\newcommand{\betatwo}{$0.09$}
\newcommand{\expansions}{\ref{tab:expansions}}
\newcommand{\betathree}{$1.0$}
\newcommand{\bb}{$\beta$}
\newcommand{\ratio}{$I_{\rm{B}}/I_{\rm{N}}$}
\newcommand{\betaratio}{$\frac{T_{e^-}}{I_{ions}}$}
\newcommand{\unit}{km s$^{-1}$}
\newcommand{\geff}{$\gamma_{\rm{eff}}$}

\newcommand{\position}{$05\rm{:}09\rm{:}31.086$ ,$-67\rm{:}31\rm{:}16.90$}%updated 3/4/14-checked 12_18
\newcommand{\traceone}{$05\rm{:}09\rm{:}28.744$ ,$-67\rm{:}31\rm{:}30.67$}
\newcommand{\tracetwo}{$05\rm{:}19\rm{:}32.038$ ,$-69\rm{:}01\rm{:}48.55$}
\newcommand{\tracethree}{$05\rm{:}19\rm{:}29.535$ ,$-69\rm{:}02\rm{:}05.31$}
\title{Constraints on Cosmic-ray Acceleration Efficiency in Balmer Shocks of Two Young Type Ia Supernova Remnants in the Large Magellanic Cloud}

\author{Luke Hovey$^1$, John P. Hughes$^{2,3}$, Curtis McCully$^{4,5}$, Viraj Pandya$^6$,and Kristoffer Eriksen$^1$}
\affil{$^1$Theoretical Design Division, Los Alamos National Laboratory, Los Alamos NM 87545, USA; lhovey@lanl.gov\\$^2$Department of Physics and Astronomy, Rutgers University, 136 Frelinghuysen Road, Piscataway, NJ 08854, USA\\   $^3$Center for Computational Astrophysics, Flatiron Institute, 162 Fifth Avenue, New York, NY 10010, USA\\  $^4$Las Cumbres Observatory, Goleta, CA 93117, USA\\  $^5$Department of Physics, University of California, Santa Barbara, CA 93106, USA\\  $^6$UCO/Lick Observatory, Department of Astronomy and Astrophysics, University of California, Santa Cruz, CA 95064, USA}

\begin{abstract}
    We present results from an optical study of two young Balmer-dominated remnants of SNIa in the Large Magellanic Cloud, \remnant{} and \remtwo{}, in an attempt to search for signatures of efficient cosmic-ray (CR) acceleration.  We combine proper motion measurements with corresponding optical spectroscopic measurements of the \ha{} line at multiple rim positions from VLT/FORS2 and SALT/RSS and compare our results to published Balmer shock models.  Analysis of the optical spectra result in broad \ha{} widths in the range of 1800--4000 km s$^{-1}$ for twelve separate Balmer-dominated filaments that show no evidence for forbidden line emission; the corresponding shock speeds from proper motion measurements from HST span a range of 1700--8500 \unit{}.  Our measured values of shock speeds and broad \ha{} widths in \remnant{} and \remtwo{} are fit well with a Balmer shock model that does not include effects of efficient CR acceleration.  We determine an upper limit of 7\%/$\chi$ (95\% confidence) on the CR acceleration efficiency for our ensemble of data points, where $\chi$ is the ionization fraction of the pre-shock gas.  The upper limits on the individual remnants are 6\%/$\chi$ (\remnant{}) and 11\%/$\chi$ (\remtwo{}). These upper limits are below the integrated CR acceleration efficiency in the Tycho supernova remnant, where the shocks predominantly show little H$\alpha$ emission, indicating that Balmer-dominated shocks are not efficient CR accelerators.
\end{abstract}

\vspace{0.4in}

\section{Introduction}
    \begin{figure*}
        \centering
        \includegraphics[scale=.7]{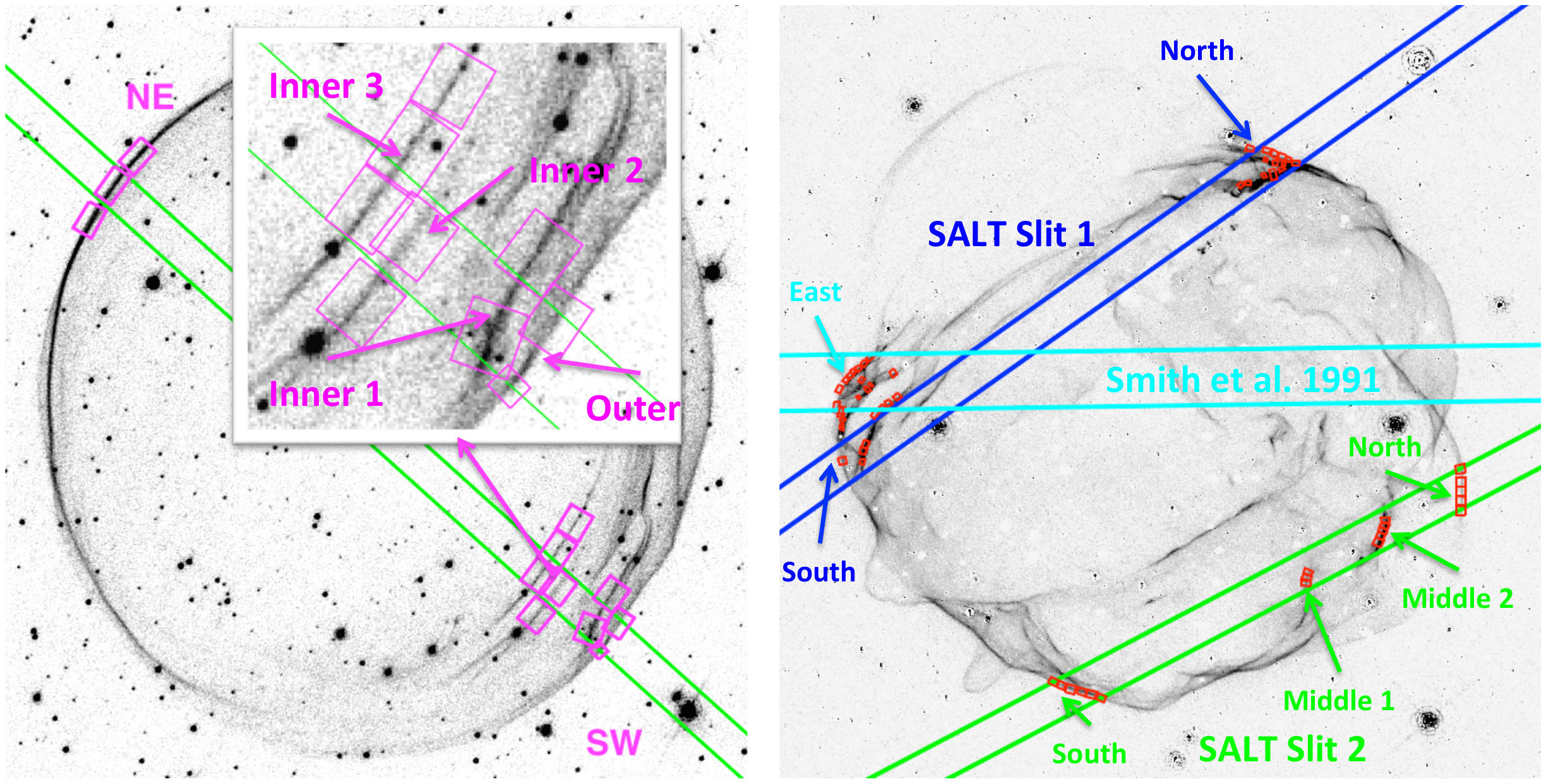}
        \caption{{\it left} - Narrow-band \ha{} image of \remnant{} taken with the ACS wide-field camera aboard \HST{} is shown.  The longslit for the FORS2 spectrum is also shown, along with a blown-up in-lay of the southwestern region of the remnants.  Proper motion measurements were made in HHE15 for all of the \ha{} extraction apertures (see table 1 and 2 therein).  All four filaments are labeled within the inlay. {\it right} - ACS narrow-band image of \remtwo{} is shown along with the longslit position (cyan) and areas of spectroscopic extraction used in \citet{smith1991}, which we label as east. We do the same with the slit positions where we obtained optical spectra with the RSS on SALT. Slit 1, shown in blue, is divided into a north and south spectral extraction regions, while slit 2, seen in green, is divided into a north, middle 2, middle1, and south spectral extraction apertures.}
        \label{fig:slits}
    \end{figure*}

    There are now multiple lines of evidence that shocks in supernova remnants (SNRs) are responsible for the acceleration of the bulk of galactic cosmic-rays (CRs) \citep{axford1981,blanford1987} up to energies of $10^{14}$ eV \citep{lagage1983}.  Observations of the Tycho SNR \citep{eriksen2011} suggest the presence of CRs at energies approaching the ``knee'' of the CR spectrum at  $\sim 10^{15}$ eV.  

    X-ray observations of many young SNRs reveal thin filaments with hard featureless spectra that are consistent with synchrotron emission from TeV energy electrons accelerated in the high speed shocks of these objects \citep[for a thorough review, see][]{reynolds2008}. 
    
    Additionally, there have been GeV and TeV $\gamma$-ray detections from RCW 86 \citep{aharonian2009},  SN 1006 \citep{acero2010}, Cas A \citep{abdo2010}, and Tycho \citep{giordano2012} SNRs, along with several others.  However, it is still unclear in some cases if the $\gamma$-ray emission results from leptonic or hadronic processes.  In the hadronic process, neutral pions are created when accelerated protons interact with the ambient material, which would be strong evidence for efficient CR acceleration in the SNR's forward shock ($\pi_0$ decay to 2$\gamma \sim 98.8\%$ of the time \citep{pdg2004}).  If instead the GeV/TeV emission process is leptonic, low energy background photons, such as photons from the cosmic microwave background, are inverse Compton scattered by relativistic electrons in the ionized plasma of the remnant.  The leptonic scenario, however, does not preclude the acceleration of CRs.  In general the leptonic emission process does not require the strong magnetic field amplification necessitated by the hadronic process, and hence requires lower acceleration efficiencies.
    
    Balmer-dominated (BD) shocks provide another diagnostic of CR acceleration.  The optical spectra of BD shocks show only Balmer series emission lines with little or no detectable forbidden emission lines.  The mean-free-path length for particle collisions in these shocks is on the scale of parsecs, hence they are referred to as {\em collisionless}, where the shocks are mediated by electromagnetic forces via plasma instabilities \citep[for a thorough review, see][]{marcowith2016}.  In this situation, ionized preshock material experiences the shock front, whereas neutral hydrogen enters the shock with no interaction.  Hydrogen lines observed in these shocks show both broad and narrow components.  The narrow component comes from the collisional excitation of neutral hydrogen atoms in the post-shock region, while the broad component comes from a postshock ion that acquires an electron through charge exchange with a neutral atom. The broad component is useful for detecting signatures of efficient CR acceleration since it indicates the temperature of post-shocked hydrogen ions. 
    
    In a J-shock for a $\gamma=5/3$ gas, the temperature of a post-shocked ion or electron is,
        \begin{equation}
            k_{\rm B} T_{i,e}=\frac{3}{16}m_{i,e} {v_{\rm s}}^2\rm{,}
        \end{equation}
    where $i,e$ is the particle species in question.  Equation 1 is valid if and only if each species thermalize their own kinetic energy according to the Rankine-Hugoniot condition and there are no plasma or collisional processes that are able to transfer energy from protons to electrons in the shock layer.  In this case the proton to electron temperature ratio is equal to the ratio of their masses.  This ratio is modified when the effects of electron heating due to ion acoustic and Langmuir waves are considered \citep{cargill1988}.  Proton temperatures, meanwhile, are sensitive to the acceleration of CRs since the kinetic energy of the forward shock serves as the reservoir for the acceleration of CRs.  Though there exists a degeneracy between the ratio of the post-shock electron to ion temperature and the efficiency at which CRs are accelerated, we can place meaningful upper bounds on the acceleration efficiency under the assumption that there is no equilibration between the temperatures of post-shock electrons and ions.  This assumption may be unrealistic, but it allows the most conservative constraint on the upper bounds of CR acceleration efficiency.

    The SNRs \remnant{} and \remtwo{} were discovered as X-ray sources by the \emph{Einstein Observatory} \citep{long1981} and confirmed as Balmer-dominated (BD) supernova remnants by \citet{tuohy1982}.  These remnants have been typed as having a Ia origin for several reasons, such as the need for a partially neutral ambient medium that would allow for the strong Balmer emission seen in these objects \citep{chevalier1980,tuohy1982}.  \citet{hughes1995} found their X-ray spectra exhibited an abundance of Si and Fe compared to oxygen group elements, which is indicative of a Ia origin.  \citet{yamaguchi2014} argued that SNRs \remnant{} and \remtwo{} are of Ia origin since the centroid of the k$\alpha$ line is at a lower ionization state, which is indicative of remnants of Ia origin compared to the core-collapse remnants that have higher ionization states.  \citet{lopez2009} and  \citet{peters2013} also argued that these remnants are Ia based upon the symmetrical morphology observed in X-ray and infrared.  SNR \remnant{} has been typed in the most convincing fashion through optical spectroscopy of its light echoes by \citet{rest2008b}, that found the original event was an over-luminous Ia supernova like SN 1991T; a conclusion supported by the X-ray properties of the SNR \citep{badenes2008}.  Both of these SNRs are young, less than 1000 years old \citep{hughes1995,rest2005lmc} and are therefore expanding at high speed ($>$1000 km s$^{-1}$).

    A great advantage of studying these remnants is that they are situated in the Large Magellanic Cloud.  This means we know the distance to these remnants with much greater accuracy than many of the Galactic remnants, which allows us to translate angular expansion rates into physical shock speeds.  We therefore can directly compare the shock speeds of these remnant filaments to the measured broad \ha{} widths to search for signs of efficient CR acceleration.  We assume a distance to the LMC of 50 kpc with an uncertainty of 4\% \citep{clementini2003}.
    
    Our paper is structured as follows.  In \S2 we detail the reductions of the data used in this analysis.  We present our proper motion measurements and spectroscopic measurements of the \ha{} line for the two remnants in \S 3.  In \S 4 we compare our measured shock speeds to the measured \ha{} widths using the Balmer shock model of \citet{morlino2013c} in order to place upper limits on CR acceleration efficiency and the degree of temperature equilibration between post-shock ions and electrons.  Our conclusions are given in \S 5.
    
    \begin{figure*}
        \centering
        \includegraphics[scale=.47]{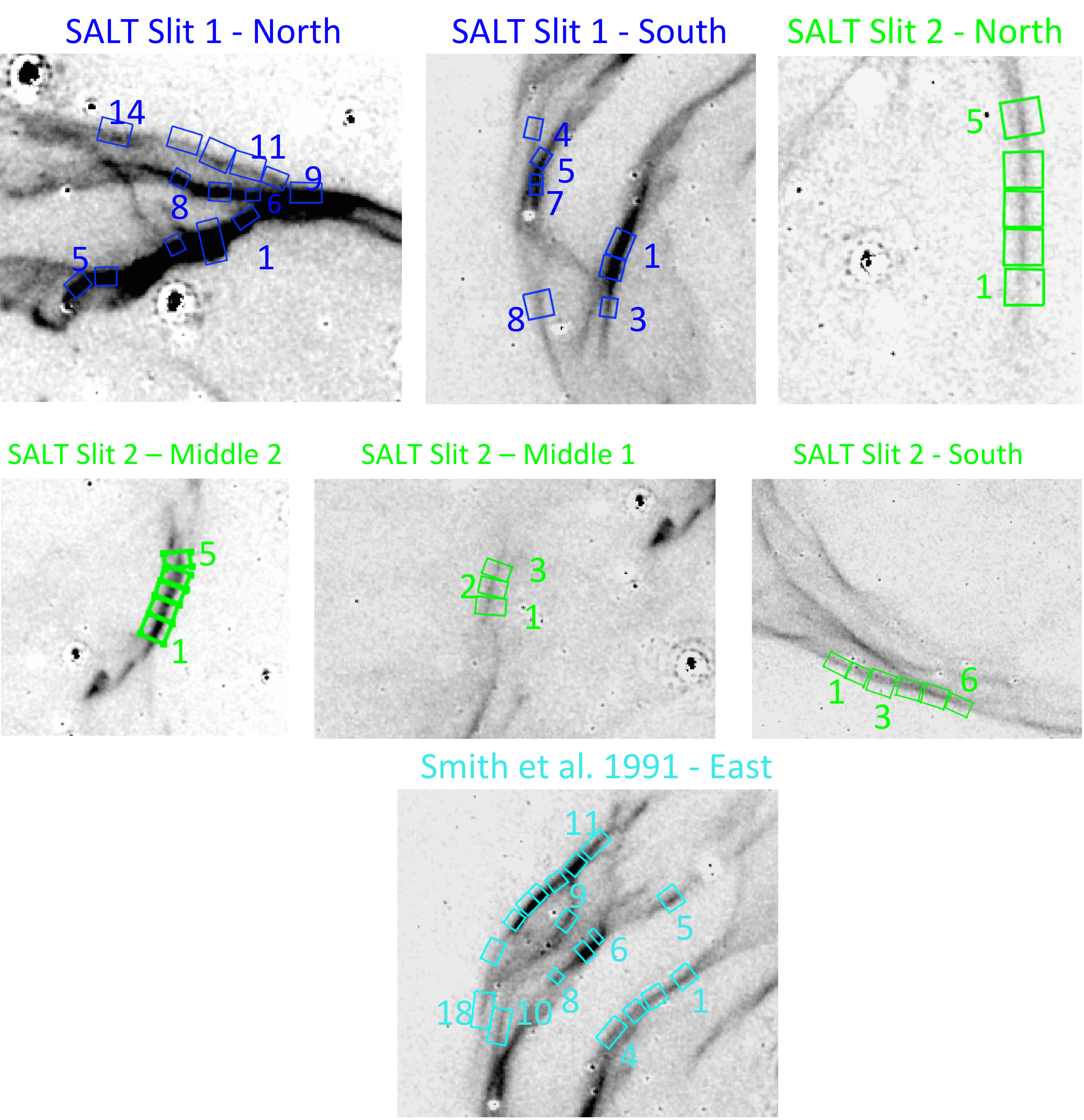}
        \caption{Zoom in of the proper motion apertures that correspond to the spectral extraction regions for which we obtained broad \ha{} widths in SNR \remtwo{} from our first epoch narrow-band \ha{} ACS image (see the right panel of figure \ref{fig:slits}.  Each panel shows the extraction regions from which we measure the proper motions in each aperture that are reported in table \expansions{}.}
        \label{fig:regions}
    \end{figure*}
    
\section{Data and Reductions}
    \subsection{HST Imaging}
        We use the two epoch \HST{} imaging in narrow-band \ha{} of \remnant{} that was presented in \citet{hovey2015}, hereafter HHE15. The observations of \remnant{} that we use here are from that work.  We use two narrow-band \ha{} \HST{} observations of SNR \remtwo{} separated by $\sim$1 year (369 days).  Our first epoch observation was imaged with the Advanced Camera for Surveys (ACS) Wide-Field (WF) camera on 2010 April 17 using the F658N \ha{} filter for a total integrated exposure time of 4757 seconds under \HST{} Program number 12107 (PI: J.P.~Hughes).  The second epoch observation was taken on 2011 April 21 with the ACS/WF camera for an integrated time of 4757 seconds using the same F658N \ha{} narrow-band filter.\footnote{Based on observations made with the NASA/ESA Hubble Space Telescope, obtained at the Space Telescope Science Institute, which is operated by the Association of Universities for Research in Astronomy, Inc., under NASA contract NAS5-26555.}
        
        Both epochs of \remtwo{} were processed using the {\tt astrodrizzle} and {\tt tweakwcs} routines, which are a part of the python {\tt drizzlepac} package.  The combined images were drizzled onto a 0.025$^{\prime\prime}$ pixel$^{-1}$ scale with a {\it pixfrac} setting of 0.8 and square convolution kernel.  The combined drizzled images were aligned using $\sim$500 field stars near the center of \remtwo{} with a positional R.M.S. uncertainty of 0.0028$^{\prime\prime}$.  We take this uncertainty in our alignment to be our systemic uncertainty.
    \subsection{Optical Spectroscopy}
        We reanalyzed the FORS2 spectrum of \remnant{} from \citet{helder2010,helder2011} (program number 384.D-0518(A), PI: E. A.~Helder). These data were taken on 2009 October 16, 2009 October 21, and 2009 November 11 with integration times of 2734 seconds, 2734 seconds, and 5468 seconds respectively.  We do not use the data obtained on 2009 October 16 since the seeing deteriorated from 1.0$^{\prime\prime}$ to 2.3$^{\prime\prime}$ during that observation.  The seeing for the other nights was $\sim$0.8$^{\prime\prime}$, which is more suited to our study of the closely spaced \ha{} filaments in the southwestern portion of \remnant{}.
        
        The FORS2 data were reduced using standard IRAF\footnote{IRAF is distributed by the National Optical Astronomy Observatory, which is operated by the Association of Universities for Research in Astronomy (AURA) under a cooperative agreement with the National Science Foundation.} tasks. We subtracted the master bias, used lamp flats to correct for inter-pixel sensitivity, and calibrated the wavelength solution using arc lamp observations. These spectra were flux calibrated using observations of the spectrophotometric standard stars HD49798 \citep{jaschek1963} and LTT2415 \citep{hamuy1992}.
        
        The spectra from SNR \remtwo{} were taken with the prime focus Robert Stobie Spectrograph (RSS) on the Southern African Large Telescope (SALT) at two different positions with a 2$^{\prime\prime}$-wide longslit and the pg0900 grating.  The slit positions are shown in figure \ref{fig:slits} {\it (right)}.  The spectrum from the position labeled  slit 1 was taken on 2011 October 03 for 1500 seconds and on 2014 March 7 for 3000 seconds.  Slit 2 was taken on 2014 March 12 for 1500 seconds and on 2014 March 19 for 1300 seconds.  The SALT observation in 2011 was taken under proposal ID 2011-3-RU-008 (PI: Hovey) and in 2014 under proposal ID 2013-2-RU-003 (PI: Hughes).
    
        Each of the four science observations were accompanied by multiple flat-field images as well as unique argon arc lamp observations, with all the long-slit spectra tuned to a grating angle of 13.62 degrees to ensure wavelength coverage of H$\alpha$. We also observed the standard B1V star Hiltner 600 \citep{stone1996,massey1988} at the same grating angle on 2014 March 7 that was used to flux calibrate all science spectra. These SALT long-slit spectroscopic data were reduced using our own custom pipeline that incorporates standard tasks from both IRAF/PyRAF as well as PySALT \citep{crawford2010}. In addition to the standard data reduction steps for long-slit spectra, we also ran a custom version of the {\tt LA-Cosmic} \footnote{\url{http://github.com/cmccully/lacosmicx}} routine to remove cosmic rays \citep{vandokkum2001} and used PySALT functions to simultaneously derive two-dimensional wavelength calibrations and rectify curved arc and telluric emission lines. The wavelength solution was further refined using the extracted one-dimensional arc lamp spectra. Standard IRAF/PyRAF functions were used to fit and subtract telluric and background emission.   
    
\begin{figure}
            \centering
            \includegraphics[scale=.42]{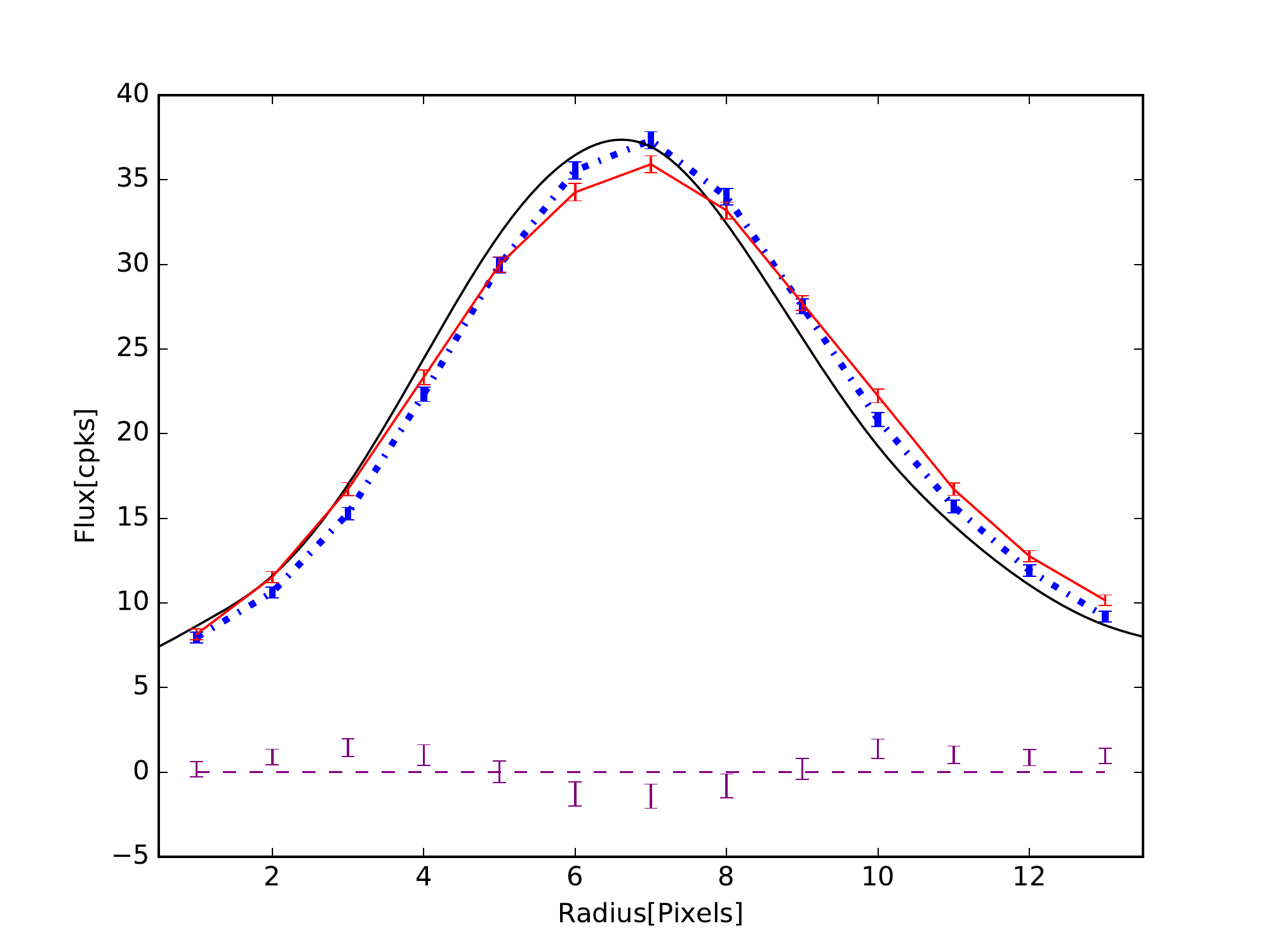}
            \caption{1-D spatial profiles of \ha{} flux, given in counts per 1$\times 10^3$ seconds (cpks), for region 1 in the southern extraction aperture for slit 1 in \remtwo{}. The black line is the \ha{} brightness profile extracted from the first epoch image and the the second epoch profile is shown in red.  A blue curve shows the shifted brightness profile of the first epoch profile to the second epoch where the modified \stat{} statistic \citep{hovey2015} is minimized.  The residuals from the subtraction of the second epoch profile with the best fit solution of the first epoch is shown in purple. }
            \label{fig:profile}
        \end{figure}
         
\section{Proper Motions and Optical Spectroscopy}

    \subsection{SNR \remnant{} Shock Speeds}
    \label{sec:0509_speeds}
        Shock speeds for the outer rim and interior filaments of SNR were measured in HHE15.  For the northeast of \remnant{}, we use the measured value of 6860$\pm$307 km s$^{-1}$, which is the weighted average of the proper motion measurements from outer regions 7, 8, and 9 from \S 4 of HHE15.  We break the southwest portion of the remnant into four distinct Balmer filaments that can be seen in the inlay of the left panel in figure \ref{fig:slits}.  The shock speeds of the outer and inner 1 filaments in the southwest are 8500$\pm$1340 km s$^{-1}$ and 3760$\pm$906 km s$^{-1}$ respectively, which are again drawn from the work done in \S 4 of HHE15.  We use the weighted arithmetic mean for inner regions 4 and 5, and then inner regions 1, 2, and 3 from HHE15 to combine the proper motions for the inner filament 2 and inner filament 3 in the southwest of \remnant{} respectively. The measured shock speed in the plane of the sky is 5700$\pm$1310 km s$^{-1}$ for the inner filament 2 and 4300$\pm$1010 km s$^{-1}$ for inner filament 3.  These results are given in table \ref{tab:results}.

    \subsection{SNR \remtwo{} Shock Speeds}
    \label{sec:0519_speeds}
        Using the same procedures as HHE15, we measured the shock speeds along various filaments in \remtwo{} for which we have corresponding spectra.  Figure \ref{fig:regions} shows the apertures from which we extracted one-dimensional \ha{} flux profiles in order to measure the proper motions.   We show an example of the \ha{} profiles from region 1 in our slit 1 southern spectral extraction aperture in figure \ref{fig:profile}.  The first epoch profile is shown in black and the second epoch profile is shown in red.  Our best fit solution for the first epoch profile shifted to the position of the second epoch is shown in blue with the corresponding residuals in purple.  In table \expansions{} we report on the measured proper motions for all filaments that lie in both of our spectroscopic slits, along with the region for which \citet{smith1991} were able to measure the broad \ha{} width. The component of the shock speed in the plane of the sky, $v_{\rm pm}$, for individual filaments are combined using the measured flux as the weighting factor as,
        \begin{equation}
            v_{\rm{pm}}=\frac{\sum\limits_{i=1}^N v_i {F_i}}
            {\sum\limits_{i=1}^N {F_i}}\rm{,}
        \end{equation}
        where $F_i$ is the integrated flux of a given 1D \ha{} brightness profile.  We choose this particular weighting due to the complicated filamentary structure since we seek to compare these shock speeds with the measured broad \ha{} widths from optical spectroscopy discussed in \S \ref{sec:saltspec}.  Shock speeds less than 400 \unit{} are excluded from the combination since they are below the spectral resolution of our observations.  These combined shock speeds are given in column 7 of table \ref{tab:results}. Note that we investigated other weighting schemes (e.g., using the inverse variance of the individual measured shock speeds), and found that our results were not sensitive to the particular choice of weighting method.  As for the quoted uncertainties, we used the larger of the  inverse variance combination of the individual filament velocity errors (for the \remtwo{} Slit 2 North region) or the RMS spread of their speeds (all other regions).

    \begin{deluxetable*}{cccccccc}
        \tablecolumns{8}
        \tablewidth{0pc} 
        \tabletypesize{\scriptsize}
        \tablecaption{Spectroscopic and Combined Proper Motion Measurements for \remnant{} and \remtwo{}}
         \tablehead{
          \colhead{Extraction Region} & 
           \colhead{FWHM \ha{} [km s$^{-1}$]} & \colhead{${\chi^2}^{\left(1\right)}$} & \colhead{d.o.f.$^{\left(1\right)}$} & \colhead{${I_{\rm B}/I_{\rm N}}^{\left(2\right)}$ } & \colhead{$v_{\rm bulk}^{\left(3\right)}$[km s$^{-1}$]}   & 
           \colhead{$v_{\rm pm}^{\left(4\right)}$[km s$^{-1}$]}    & 
           \colhead{$v_{\rm s}^{\left(5\right)}$[km s$^{-1}$]}  
           }
          \startdata
             
0509$-$67.5 NE & 4000$\pm$235 & 506 & 470 & 0.060$\pm$0.011 & 89.0$\pm$96.1 & 6900$\pm$307 & 6900$\pm$322 \\
0509$-$67.5 SW Outer & 4000$\pm$167 & 541 & 470 & 0.200$\pm$0.033 & 440$\pm$70.7 & 8500$\pm$1340 & 8500$\pm$1340 \\
0509$-$67.5 SW Inner 1 & 2900$\pm$40.6 & 407 & 470 & 0.370$\pm$0.027 & 770$\pm$21.4 & 3800$\pm$906 & 3900$\pm$906 \\
0509$-$67.5 SW Inner 2 & 3500$\pm$117 & 569 & 470 & 0.200$\pm$0.025 & 820$\pm$55.1 & 5700$\pm$1310 & 5800$\pm$1310 \\
0509$-$67.5 SW Inner 3 & 3900$\pm$136 & 642 & 470 & 0.200$\pm$0.023 & 1000$\pm$65.4 & 4300$\pm$1010 & 4500$\pm$1010 \\
0519$-$69.0 Slit 1 North & 1800$\pm$26.0 & 359 & 312 & 0.700$\pm$0.032 & -71.0$\pm$9.60 & 1900$\pm$254 & 1900$\pm$254 \\
0519$-$69.0 Slit 1 South & 2300$\pm$27.6 & 309 & 312 & 0.600$\pm$0.022 & -57.0$\pm$10.8 & 2500$\pm$886 & 2500$\pm$886 \\
0519$-$69.0 Slit 2 North & 2800$\pm$461 & 335 & 315 & 0.300$\pm$0.076 & 370$\pm$188 & 3300$\pm$990 & 3300$\pm$1010 \\
0519$-$69.0 Slit 2 Middle 2 & 1900$\pm$51.1 & 347 & 315 & 0.800$\pm$0.043 & -140$\pm$20.3 & 1700$\pm$384 & 1700$\pm$385 \\
0519$-$69.0 Slit 2 Middle 1 & 2700$\pm$87.1 & 294 & 315 & 0.700$\pm$0.029 & -240$\pm$35.2 & 3700$\pm$1070 & 3700$\pm$1070 \\
0519$-$69.0 Slit 2 South & 2700$\pm$131 & 282 & 315 & 0.400$\pm$0.027 & 190$\pm$50.5 & 2800$\pm$493 & 2800$\pm$496 \\
0519$-$69.0 Smith `91 East & ${1300\pm200}^{\left(6\right)}$ & N/A & N/A & ${0.8\pm0.2}^{\left(6\right)}$ & N/A & 1300$\pm$935 & ${1300\pm935}^{\left(7\right)}$ \\

         \enddata
         \vspace{-0.3cm} \tablecomments{$^{\left(1\right)}$The measured $\chi^2$ and corresponding degrees of freedom (d.o.f.) refer to the fit of the broad \ha{} width only.\\$^{\left(2\right)}$these values should only be treated as lower limits given the complicated shock structures where the blended narrow emission from fast and slow moving shocks down-weight these ratios. \\$^{\left(3\right)}$ Bulk velocities are in the line of sight and are obtained by the difference of the broad and narrow \ha{} centers.  \\$^{\left(4\right)}$Values are the combined proper motions for the various spatial \ha{} apertures discussed in \S \ref{sec:0509_speeds} and \S \ref{sec:0519_speeds} and assuming an LMC distance of 50 kpc. \\$^{\left(5\right)}$This is the root-sum-square of $v_{\rm pm}$ and $\left(\left(\gamma+1\right) v_{\rm bulk}/2\right)$, where $\gamma=5/3$.\\$^{\left(6\right)}$Values are from \citet{smith1991}.\\$^{\left(7\right)}$Since \citet{smith1991} did not report the offset between the broad and narrow \ha{} lines for the eastern filament, we assume that $v_{bulk}=0$ \unit{}, which is consistent with the fact that the $v_s$ values are within the $1\sigma$ uncertainties of the $v_{\rm pm}$ values for the rest of SNR \remtwo{}.
         }
        \label{tab:results}
    \end{deluxetable*}
     \begin{figure}[b]
        \centering
        \includegraphics[width=3.2truein]{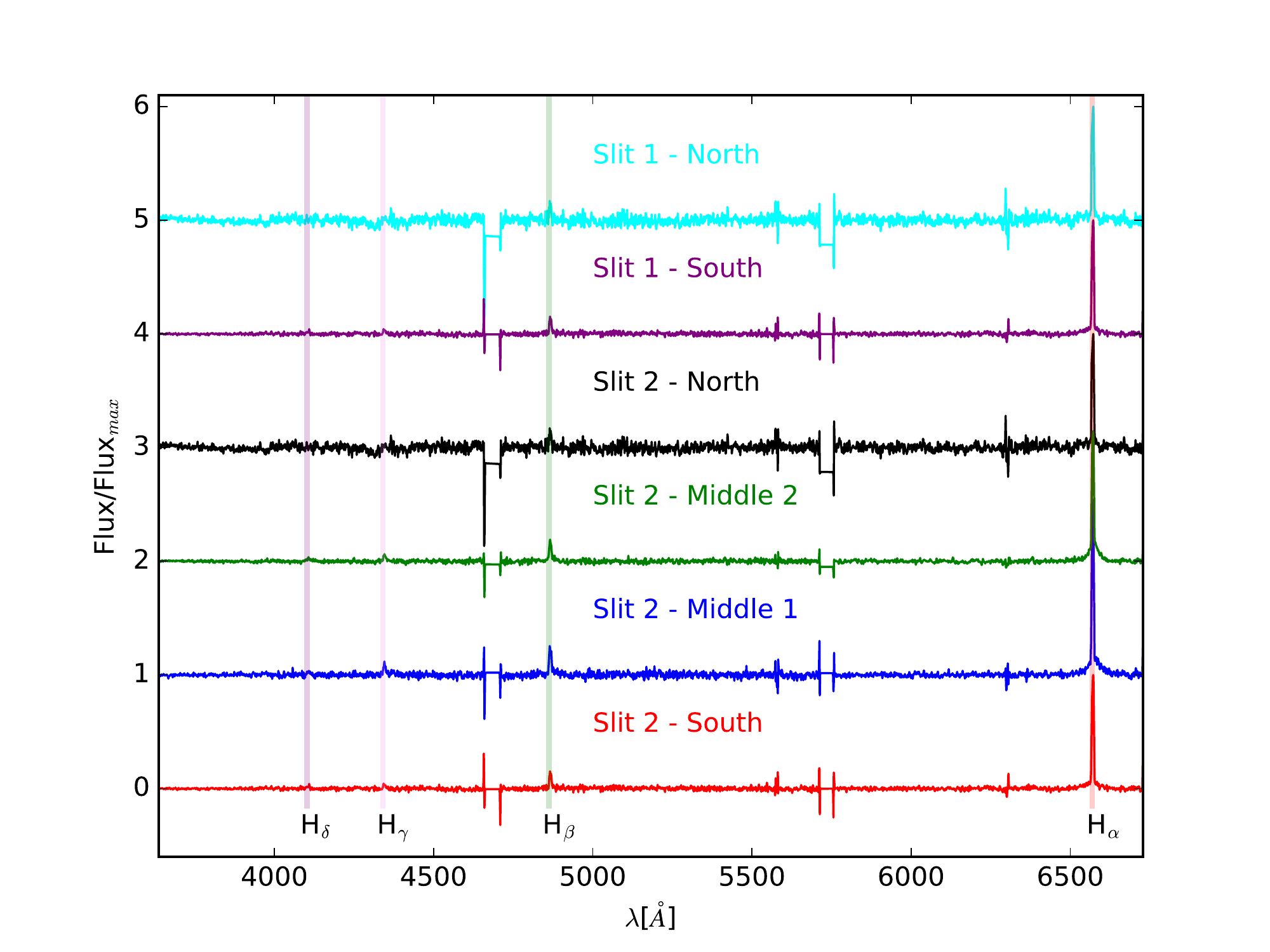}
        \caption{SALT RSS spectra for our two extraction apertures from slit 1 and four from slit 2 in SNR \remtwo{}.  The fluxes are separated by adding a factor of unity to each subsequent spectrum.}
        \label{fig:0519_spec}
    \end{figure}
    
    \begin{figure*}[t]
        \centering
        \includegraphics[scale=1.2]{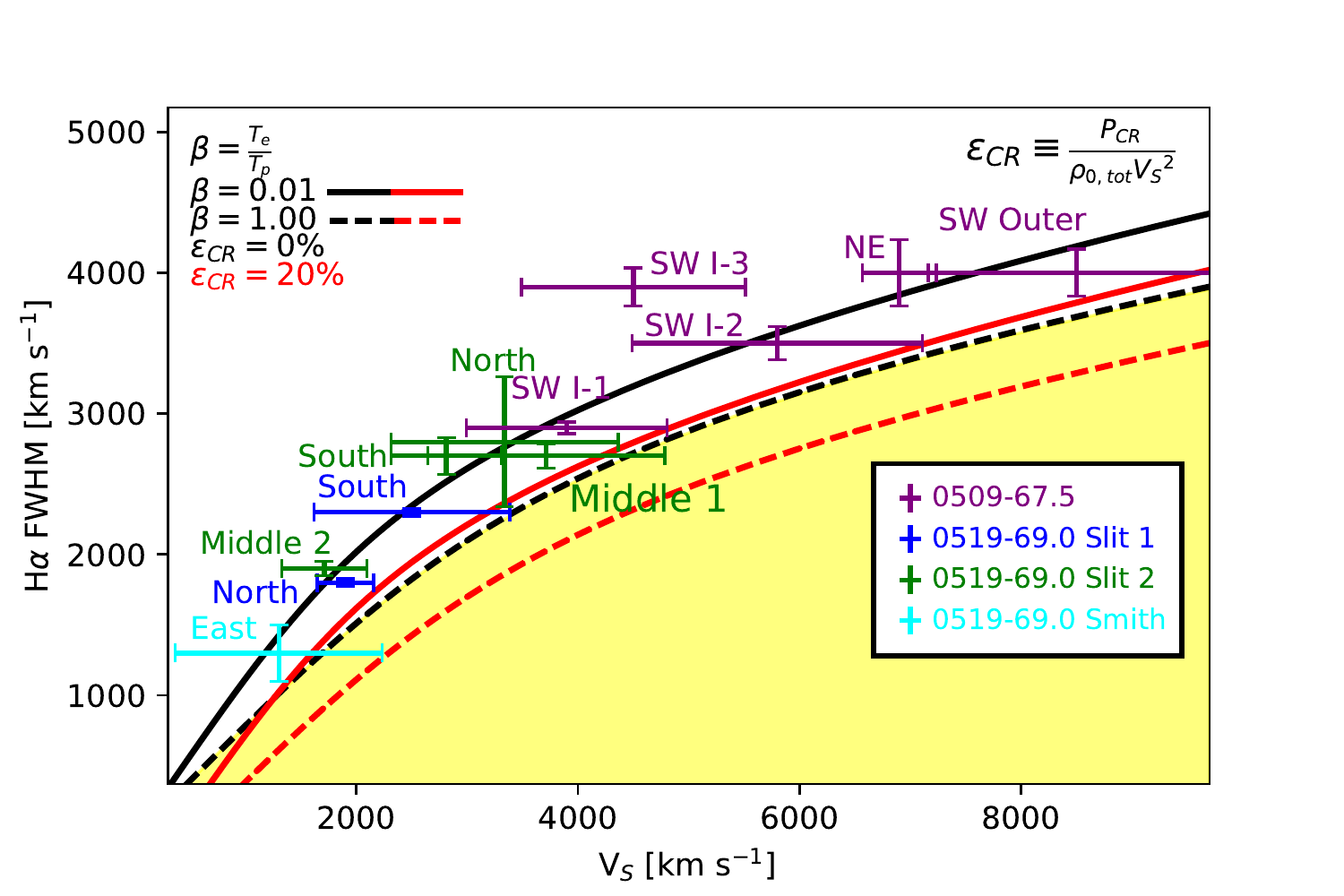}
        \caption{Comparison of the measure \ha{} line widths with corresponding shock velocities for \remnant{} and \remtwo{} with the Balmer shock model from \citet{morlino2013c}.  The solid and dashed curves represent the case of no equilibration between the temperature of post shocked electrons and ions and the case of full equilibration respectively.  The black curves are for the case of no CR acceleration ($\epsilon_{CR}=0\%$) and the red curves are for the case of a CR acceleration efficiency of $\epsilon_{CR}=20\%$.  The yellow region below the curves indicates the parameter space where a minimum CR acceleration efficiency can be determined \citep{morlino2013c}.}
        \label{fig:beta}
    \end{figure*}  
    
    \subsection{SNR \remnant{} Optical Spectroscopy}
        Our goal in the reanalysis of the FORS2 spectrum is to measure the broad \ha{} widths for the four distinct filaments in the southwest, which HHE15 argued were confused by this complicated multiple filamentary structure.  To compensate for these closely spaced shocks, we choose a spectral extraction aperture of 1$^{\prime\prime}$, which is slightly larger than the full width half maximum (FWHM) of the average seeing of the observations and corresponds to the minimum separation between inner filaments 2 and 3.  In the northeast we did not have this constraint so we use an extraction aperture that is 1.5$^{\prime\prime}$ in size to maximize the signal from the faint flux of the broad \ha{} component.   
        
        Standard IRAF procedures were employed to subtract night skylines and stellar continuum flux of the two-dimensional spectra.  A 1D spectrum was made for each of the three observations by summing the flux along the spatial extraction region for each wavelength bin.  The three resulting 1D spectra were then median-combined to produce the final spectrum for analysis.
        
        \citet{smith1994} measured the widths of the narrow \ha{} line widths in this remnant, which has a weighted mean of 27.50$\pm2.02$ \unit{}.  We use this value as the fixed FWHM of the narrow component when fitting our \ha{} spectrum in the NE and SW of \remnant{}.  The spectra are analyzed by first fitting the broad component of the \ha{} line by fitting a Gaussian, excluding the narrow component (6555-6570 \AA).  We give the values of the \stat{} statistic and the corresponding degrees of freedom (d.o.f.) in columns three and four of table \ref{tab:results}.  We then subtracted the broad component and fit the narrow component with a Gaussian convolved with a top-hat kernel with a width of $\sim$10 \AA\ that corresponds to the 1.6$^{\prime\prime}$ width of the slit.  We find that the narrow \ha{} component is centered at a wavelength of 6567.5$\pm 0.05$ \AA, which corresponds to a velocity of 213.0$\pm 1.8$ \unit{}. This is roughly consistent with the measured redshift velocity of the LMC, 262 \unit{} \citep{mcconnachie2012}.     
        
        In the northeast of \remnant{} we measured the broad \ha{} width to be 4000$\pm 235$ km s$^{-1}$, which has an uncertainty that is $\sim4$ times smaller than that reported in \citet{helder2010,helder2011}.  We report our measured values for the broad \ha{} width, broad-to-narrow ratios, and line of sight velocities (determined by the difference of the broad and narrow \ha{} line centroids) in table \ref{tab:results} and our spectra in figure \ref{fig:0509_spectra}.  Table \ref{tab:results} also gives the combined proper motion measurements for the filaments and the combined line of sight velocity with the radial speed for the full velocity of the filament.  Assuming the shocked gas has an equation of state of $\gamma=$5/3 \citep{hovey2015}, we boost the bulk velocities by a factor of 4/3, which we then combine with the proper motions to calculate the total velocities reported in table \ref{tab:results}.  
        
        In the southwest, we measured broad line widths for the outer rim and three interior filaments.  For the outer filament, we measure a broad \ha{} width of 4000$\pm 167$ km s$^{-1}$, though this should be considered a lower limit due to contamination of the spectrum from the brighter interior filament 1.  This is due to the fact that inner fillament 1 is moving at a considerably slower speed than the outer filament, which we would expect to a have a brighter broad \ha{} component.  The aperture for the outer filament is placed as far as possible from inner filament 1, but we still expect contamination that would artificially lower the measured width of the fainter, faster moving outer shock.  Even though this shock speed should be treated as a lower limit, the value we quote is conservative for the purposes of this study since larger values would only lead to a smaller calculated CR acceleration efficiency than the value which is presented in \S 4.  
        
        We measure broad \ha{} widths of 2900$\pm 40.6$ km s$^{-1}$, 3500$\pm 117$ km s$^{-1}$, and 3900$\pm 136$ km s$^{-1}$ for interior filaments 1, 2, and 3 respectively.  Again we quote our measured spectroscopic values in table \ref{tab:results} and show the spectra in figure \ref{fig:0509_spectra}.  The line of sight bulk velocities increase from the outer region progressively to the inner region 3 in the southwest with values of 440$\pm$70.7 km s$^{-1}$, 770$\pm$21.4 km s$^{-1}$, 820$\pm$55.1 km s$^{-1}$, and 1000$\pm$65.4 km s$^{-1}$ for the four filaments.

    \subsection{SNR \remtwo{} Optical Spectroscopy}
    \label{sec:saltspec}

           Slit 1 had three separate exposures and slit 2 had two.  The three 2D spectral files for slit 1 were median combined before extracting the final 1D spectrum.  For slit 2, the two 2D spectral files were averaged before extracting the 1D spectrum. 
           
           We use a fixed narrow \ha{} line width of 40.30$\pm$0.24 \unit{} in our \ha{} line fitting, which is the statistical mean of the measurements from \citet{smith1994}.  The broad \ha{} line is first fit by excluding the narrow \ha{} flux. Our \stat{} values and corresponding d.o.f. are reported in columns three and four respectively in table \ref{tab:results}.  After this the broad \ha{} line flux is subtracted from the spectrum and we fit the narrow line convolved with a Gaussian, which fit the data better than a top-hat kernel.  For both of our longslits, we find that the narrow \ha{} component is centered at a wavelength of $6569.00\pm0.08$ \AA, which translates into a redshift velocity of 283$\pm 3.7$ \unit{}.  Again this is consistent with the redshift velocity of the LMC reported by \citet{mcconnachie2012}.     
           
           Our spectra for both slits are shown in table \ref{fig:0519_spectra} and the measured \ha{} broad line widths, broad-to-narrow intensity ratios, and line of sight velocities are reported in table \ref{tab:results}.  We measure broad \ha{} widths of 1800$\pm 26.0$ km s$^{-1}$ and 2300$\pm 27.6$ km s$^{-1}$ for the north and south portions of slit 1 respectively.  We divided slit two into four regions: the north, middle 1, middle 2, and south for which we measure broad \ha{} line widths of 2800$\pm 461$ km s$^{-1}$, 1900$\pm 51.1$ km s$^{-1}$, 2700$\pm 87.1$ km s$^{-1}$, and 2700$\pm 131$ km s$^{-1}$ respectively.

\section{Discussion}    
\subsection{Comparing Results with Balmer Shock Models}
    In order to place limits on the efficiency of CR acceleration in these shocks we compare our results with the Balmer shock model from \citet{morlino2013c}.  We have chosen to only consider this model in our current analysis, which contrasts with the approach found in \S 4 of HHE15, which additionally considered the Balmer shock model of \citet{vanadelsberg2008} for the comparison of the shock speeds they measured in SNR \remnant{} and the spectroscopic measurements of broad line \ha{} widths from \citet{helder2010,helder2011}.  The decision to only employ the Balmer shock model found in \citet{morlino2013c} is based upon the following reasons:
    \begin{enumerate}
        \item There is no consideration of efficient CR acceleration or the effects that its precursor might have on the Blamer shock model of \citet{vanadelsberg2008}.  Furthermore, this model lacks a prescription for the attenuation of broad \ha{} widths at a given shock speed as CR acceleration increases, like that found in \citet{morlino2013c}.  
        \item The \citet{vanadelsberg2008} model lacks an independent treatment of material upstream from the shock, but rather assume that it is in an unadulterated thermalized state, which they cite as a weakness that they plan on addressing in the future in their concluding remarks. \citet{morlino2013c} treats the upstream material separately and found that even in the case of no CR acceleration, the upstream material is affected by the neutral return flux.
        \item \citet{vanadelsberg2008} also assumes that the populations of post-shock ions and neutrals equilibrate after five charge exchanges, and that both populations will then have the same bulk speed.  One of the main results from the \citet{morlino2013c} is that the neutrals do not thermalize and have both a lower temperature and bulk speed than the corresponding population of ions \citep{morlino2013}.  Again, this is explicitly cited as a weakness in the closing paragraphs of \citet{vanadelsberg2008}.  
    \end{enumerate}
    One potential weakness of the Balmer shock model of \citet{morlino2013c} is their exclusion of helium and its possible effects on the resulting Balmer emission. \citet{morlino2013c} explicitly address this issue and conclude that their omission of helium does not explain the  differences between their model and the one presented in \citet{vanadelsberg2008} and that in general, helium will have negligible effects on the system.  However, it is not clear what role helium may play on attenuating \ha{} broad line widths once a non-negligible CR acceleration efficiency is included.  Given the preponderance of strengths to the sole weakness of the Balmer shock calculations of \citet{morlino2013c}, we only consider this model in the interpretation of our measurements.  
    
    Figure \ref{fig:beta} shows the Balmer shock model predictions for broad \ha{} line widths as  a function of shock speed for the full range of possible post-shock electron to ion temperature ratios.  Black curves are for the case where there is no CR acceleration, while the red curves are for the case of a CR acceleration efficiency of $\epsilon_{CR}=20\%$, where $\epsilon_{CR}$ is defined in \citet{morlino2013c} to be,
    \begin{equation}
        \epsilon_{\rm CR}=\frac{P_{\rm CR}}{\rho_{\rm 0,tot}{v_{\rm s}}^2}\rm{.}
    \end{equation}
    The parameter space below the \bb{}=1 curve (shaded yellow) requires CR acceleration (or some other process draining energy from the thermal population).  Since all of our data points lie above the line, none of the points require efficient CR acceleration in the limiting case of complete temperature equilibration.
    
    In order to place an upper bound for CR acceleration efficiency we fit the ensemble of filaments to the \citet{morlino2013c} models including the effects of CRs using the conservation assumption that there is no temperature equilibration between post-shock electrons and ions ($\beta = 0.01$). We implement a Monte Carlo procedure where we perturb the shock speeds and \ha{} broad line widths within their respective uncertainties using a Gaussian random variable, running $10^5$ realizations.  We find that the CR acceleration efficiency, ${\epsilon}_{\rm CR}$, must be below \effall{} at 95\% confidence for the full ensemble of filaments.  
    
    We have also determined 95\% confidence upper bounds for the CR acceleration efficiencies for each remnant separately, and find these to be \effrem{} and \effremtwo{} for \remnant{} and \remtwo{}, respectively.
 
     Using this same Monte Carlo approach, we determine the upper bounds for the CR acceleration efficiencies for the individual filaments again for the case of no temperature equilibration between post-shock ions and electrons (Table \ref{tab:eff_results}). The individual $\epsilon_{\rm CR}$ values vary broadly from nearly 0\% to 66\%, with an average of 29\%.  The largest upper limit value comes from the eastern filament studied by \citet{smith1991}.  This region has a wide range of shock speeds from the multiple internal features it contains (see Figure~\ref{fig:speeds.in.regions}) which explains its relatively large shock velocity uncertainty and, by extension, its high upper limit on $\epsilon_{\rm CR}$.

    \subsection{Comparison to Tycho's SNR}
     We compare these results to those of Tycho's supernova remnant, which has a similarly young age, 445 years, as the LMC remnants and also exhibits BD shocks (albeit over only a very small portion of the rim).  Using their CR-hydro NEI model constrained by the structure and broadband spectrum of Tycho's SNR, \citet{slane2014,slane2015} found that $16\%$ of the kinetic energy in the remnant has been converted into relativistic particles, with $11\%$, of those escaping as CRs.  They find the present diffuse shock acceleration (DSA) efficiency in Tycho's SNR to be $26\%$.  The DSA CR efficiency is equivalent to the ion acceleration efficiency, $\epsilon^*_{\rm CR}$, from \citet{morlino2013c} defined as,
     \begin{equation}
     \epsilon^*_{\rm CR}=\frac{P_{\rm CR}}{\rho_{\rm 0,ion}{v_{\rm s}}^2}\equiv\frac{\epsilon_{\rm CR}}{{\chi}}\rm{,}    
     \end{equation}
     where $\chi$ is the ionization fraction of the pre-shock medium. In HHE15 we constrained this ionization fraction to be in the range 0.4-0.7 for plausible evolutionary models that mimic a range of CR acceleration efficiencies at the forward shock through the use of an effective equation of state. In the following, therefore, we use an ionization fraction of $1/2$, for which the ion acceleration efficiency becomes $\epsilon^*_{\rm CR}=2\epsilon_{\rm CR}$. This gives us upper limits at 95\% confidence of $\epsilon^*_{\rm CR}=12\%$ for SNR \remnant{}, $\epsilon^*_{\rm CR}=22\%$ for SNR \remtwo{}, and $\epsilon^*_{\rm CR}=14\%$ using the ensemble of all data points. 
     
    It is apparent that neither \remnant{} nor \remtwo{} is as efficient as Tycho's SNR at accelerating CRs.  As mentioned above, the BD shocks in Tycho's SNR cover only a very small fraction of the remnant (mostly in the northeastern region of the rim), while such shocks cover the entire extent of both LMC remnants. This difference could be the key to the difference in the CR acceleration efficiencies. BD shocks require a partially neutral upstream medium where processes such as ion-neutral wave damping could suppress CR acceleration if the ambient magnetic field of the upstream material is sufficiently weak \citep{drury1996}.
    
    Another plausible reason for the brighter H$\alpha$ emission {\it and} smaller CR acceleration efficiencies in SNRs 0509$-$67.5 and 0519$-$69.0 compared to Tycho's SNR, is that a stronger CR precursor exists in Tycho that is ionizing the upstream ambient medium to a greater degree, hence suppressing H$\alpha$ emission in the remnant.  \citet{knezevic2017} found a global H$\alpha$ narrow-line width in the northeastern region of $54.8\pm1.8$ km ${\rm s}^{{-1 }}$ for Tycho's SNR, whereas \citet{smith1994} measured narrow line widths of $25-31$ km s$^{-1}$ and $39-42$ km s$^{-1}$ for SNRs 0509$-$67.5 and 0519$-$69.0 respectively.  The relatively smaller narrow-line widths for 0509$-$67.5 and 0519$-$69.0 suggests that a stronger CR precursor is likely for Tycho's SNR compared to the LMC remnants we have studied.  
    
\begin{deluxetable}{ccc}
        \tablecolumns{3}
        \tablewidth{0pc} 
        \tabletypesize{\scriptsize}
        \tablecaption{CR Acceleration Efficiency Limits and Temperature Equilibration Ratios for \remnant{} and \remtwo{}}
         \tablehead{
          \colhead{Extraction Region} & 
           \colhead{${\epsilon_{\rm CR;upper}}^{\left(1\right)}$} & 
           \colhead{${\beta}_{upper}$ $^{\left(2\right)}$}
           }
          \startdata
0509$-$67.5 NE & 0.13 & $0.42$ \\
0509$-$67.5 SW Outer & 0.29 & ... \\
0509$-$67.5 SW Inner 1 & 0.28 & ... \\
0509$-$67.5 SW Inner 2 & 0.33 & ... \\
0509$-$67.5 SW Inner 3 & 0.00 & ... \\
0519$-$69.0 Slit 1 North & 0.21 & $0.84$ \\
0519$-$69.0 Slit 1 South & 0.35 & ... \\
0519$-$69.0 Slit 2 North & 0.46 & ... \\
0519$-$69.0 Slit 2 Middle 2 & 0.19 & $0.56$ \\
0519$-$69.0 Slit 2 Middle 1 & 0.41 & ... \\
0519$-$69.0 Slit 2 South & 0.13 & $0.38$ \\
0519$-$69.0 Smith `91 East & 0.66 & ... \\
\hline \\
\vspace{-15 pt} \\
$0509-67.5$ & 0.06 & 0.47 \\
$0519-69.0$ & 0.11 & 0.55 \\
All Points & 0.07 & 0.25 \\
\vspace{-17 pt} \\

        \enddata
        \tablecomments{
        ${\left(1\right)}$ - Upper limits at 95\% confidence for CR acceleration efficiency assuming no equilibration between electron and ion temperatures ($\beta=0.01$).\\
        ${\left(2\right)}$ - Upper-limit values at 95\% confidence for $\beta$, unless it cannot be unconstrained between the limits of $0.01\leq\beta\leq1$.}
        \label{tab:eff_results}
    \end{deluxetable}

    \subsection{Temperature Equilibration}
     We can use the Balmer shock models of \citet{morlino2013c} to place constraints on the degree to which the temperatures of post-shock electrons and ions have equilibrated under the assumption of no CR acceleration.  The temperature ratio of post-shock electrons to ions, $\beta$, for the several regions are given in table \ref{tab:eff_results}. Many of the regions we are unable to constrain the value of $\beta$.  For the ensemble of all data points we find an upper limit to the temperature ratio of $\beta\leq0.25$ at 95\% confidence. This upper limit is consistent with the $\beta \sim v_{\rm s}^{-2}$ relationship between shock velocity and $\beta$ established by \citet{ghavamian2007b}.

\section{Conclusions}

Using archival VLT FORS2 spectroscopy for SNR \remnant{} and SALT RSS spectra of \remtwo{} we have measured broad \ha{} widths for 12 separate regions of the forward shock of these remnants and find a range of widths of 1800--4000 \unit{}. The corresponding shock speeds of these Balmer filaments from proper motion using narrow-band \ha{} imaging with \HST\ are 1700--8500 \unit{}.  Using the offset between the broad and narrow component centroids of the \ha{} emission lines, we find bulk speeds of the material behind the shock to be in the range of 57--1000 \unit{}.  The shock speed of the eastern filament that \cite{smith1991} measured the broad \ha{} width of 1300$\pm200$ \unit{} is also measured in this study, where we find a speed of 1700$\pm953$ \unit{}. The final 3D shock velocities for the 12 newly measured filaments are determined by root-sum-square combining of the proper motions, which measure the speed in the plane of the sky, and the bulk speeds of the shocked material.

The 3D shock velocities and broad \ha{} widths are then compared to the Balmer shock model from \citet{morlino2013c}, where we find the majority of our data points are fit well with the model that did not include the effects of CR acceleration or CR precursors on the upstream material.  Using this model, we find that the temperature equilibration ratio of the post-shock electrons and ions, $\beta$, to be less than 25\% for our ensemble of data points at 95\% confidence.    

We use the prescription for modifying the Balmer shock models to include the effects of efficient CR acceleration from \citet{morlino2013c} and find that at 95\% confidence the maximum ion efficiency for the ensemble of data points for both remnants to be 7\%/$\chi$, where $\chi$ is the ionization fraction of the upstream material.  The 95\% confidence limits for the maximum efficiency for the individual remnants are found to be 6\%/$\chi$ and 11\%/$\chi$ for \remnant{} and \remtwo{} respectively.  At this time it is impossible to determine lower bounds for the CR efficiencies since we are unable to draw significant constraints for the temperature equilibration ratio of the post-shock electrons and ions, $\beta$, is not known and the points do fall in a region of parameter space where the efficiency and $\beta{}$ are degenerate.   

Our results are then compared to Tycho's SNR, which is also a young Ia that exhibits limited regions with BD shocks.  Besides the age similarity, this remnant is a useful comparison object for SNR \remnant{} and SNR \remtwo{} since GeV/TeV $\gamma$-rays, indicative of CR production, have been detected from Tycho's SNR \citep{giordano2012}.  Using hydrodynamic modeling, \citet{slane2014,slane2015} found that an ion acceleration efficiency of 26\% best fit the data for Tycho's SNR.  This value is higher than our upper limit on the efficiency for both LMC SNRs when we adopt an ionization fraction of 0.5 for the upstream material and it exceeds the efficiency we find for the complete ensemble of our data points (14\% including the same ionization fraction).  These results strongly suggest that BD shocks are regions of low CR acceleration, which is consistent with the complete shell of \ha{} emission around the entirety of both of the LMC remnants that we consider.  

\section{Acknowledgements}

L.H.\ would like to thank Pat Slane for helpful discussions regarding the CR efficiency of Tycho's SNR.  Some of the observations reported in this paper were obtained with the Southern African Large Telescope (SALT). Funding for SALT is provided in part by Rutgers University, a founding member of the SALT consortium. This article was partially supported by NASA/STScI grants numbered HST-GO-11015.01-A (SNR 0509$-$67.5) and HST-GO-12017.01-A (SNR 0519$-$69.0).  J.P.H.\ acknowledges the hospitality of the Flatiron Institute, which is supported by the Simons Foundation.  C.M.\ is supported by NSF grant AST-1313484.  V.P.\ is grateful to Saurabh Jha and Rachel Somerville for encouragement and support while working on the data reduction at Rutgers. We thank John Raymond and the anonymous referee for their helpful comments on the manuscript.

\clearpage
\appendix
\FloatBarrier
\section{Proper Motions for \remtwo{}}
\setcounter{table}{0}
\setcounter{figure}{0}

\renewcommand\thetable{A-\arabic{table}}
\renewcommand\thefigure{A-\arabic{figure}}
    
\begin{deluxetable}{cccccc}[H]
        \tablecolumns{6}
        \tablewidth{0pc} 
        \tabletypesize{\scriptsize}
        \tablecaption{Proper Motion Measurements for SNR \remtwo{}}
        \tablehead{
        \colhead{Slit Position} & \colhead{ID} & \colhead{Shift [mas]} & \colhead{$\chi^2$ } & \colhead{d.o.f.} & \colhead{$v_{pm}$ [km/s]$^{^{\left(1\right)}}$}
       }
        \startdata
Slit 1 & North-1    & 5.80$\pm$1.25     & 22  & 6     & 1400$\pm$305 \\
Slit 1 & North-2    & 7.80$\pm$0.75     & 77  & 21    & 1900$\pm$205 \\
Slit 1 & North-3    & 6.50$\pm$2.50     & 77  & 8     & 1600$\pm$621 \\
Slit 1 & North-4    & 8.50$\pm$1.75     & 7.9 & 8     & 2000$\pm$412 \\
Slit 1 & North-5    & 7.80$\pm$2.00     & 13  & 8     & 1900$\pm$489 \\
Slit 1 & North-6    & 8.30$\pm$3.00     & 14  & 5     & 2000$\pm$708 \\
Slit 1 & North-7    & 8.30$\pm$1.50     & 20  & 9     & 2000$\pm$342 \\
Slit 1 & North-8    & 11.0$\pm$1.75     & 7.8 & 7     & 2600$\pm$443 \\
Slit 1 & North-9    & 8.50$\pm$1.00     & 87  & 9     & 2000$\pm$257 \\
Slit 1 & North-10   & 8.50$\pm$3.75     & 29  & 7     & 2000$\pm$905 \\
Slit 1 & North-11   & 8.50$\pm$3.25     & 53  & 11    & 2000$\pm$806 \\
Slit 1 & North-12   & 7.30$\pm$3.75     & 35  & 13    & 1700$\pm$889 \\
Slit 1 & North-13   & 8.00$\pm$8.50     & 26  & 10    & 1900$\pm$2000 \\
Slit 1 & North-14   & 8.80$\pm$2.50     & 27  & 11    & 2100$\pm$621 \\
Slit 1 & South-1    & 7.30$\pm$0.75     & 60  & 12    & 1700$\pm$198 \\
Slit 1 & South-2    & 12.0$\pm$1.25     & 58  & 13    & 2900$\pm$274 \\
Slit 1 & South-3    & 9.80$\pm$2.00     & 15  & 9     & 2300$\pm$459 \\
Slit 1 & South-4    & 13.0$\pm$2.50     & 10  & 9     & 3200$\pm$584 \\
Slit 1 & South-5    & 6.30$\pm$2.00     & 4.7 & 8     & 1500$\pm$497 \\
Slit 1 & South-6    & 6.80$\pm$4.50     & 14  & 6     & 1600$\pm$1050 \\
Slit 1 & South-7    & 17.0$\pm$3.25     & 20  & 6     & 3900$\pm$798 \\
Slit 1 & South-8    & 17.0$\pm$5.25     & 25  & 15    & 4000$\pm$1280 \\
Slit 2 & North-1    & 14.0$\pm$3.50     & 29  & 19    & 3300$\pm$851 \\
Slit 2 & North-2    & 13.0$\pm$6.25     & 53  & 19    & 3200$\pm$1470 \\
Slit 2 & North-3    & 6.50$\pm$5.00     & 58  & 19    & 1600$\pm$1170 \\
Slit 2 & North-4    & 20.0$\pm$10.8     & 29  & 19    & 4800$\pm$2550 \\
Slit 2 & North-5    & 14.0$\pm$2.75     & 49  & 19    & 3400$\pm$638 \\
Slit 2 & Middle 2-1 & 6.30$\pm$1.00     & 42  & 17    & 1500$\pm$223 \\
Slit 2 & Middle 2-2 & 7.50$\pm$1.25     & 35  & 17    & 1800$\pm$282 \\
Slit 2 & Middle 2-3 & 6.00$\pm$1.50     & 360 & 17    & 1400$\pm$359 \\
Slit 2 & Middle 2-4 & 9.80$\pm$1.75     & 16  & 17    & 2300$\pm$412 \\
Slit 2 & Middle 2-5 & 5.00$\pm$2.50     & 51  & 17    & 1200$\pm$597 \\
Slit 2 & Middle 1-1 & 16.0$\pm$8.75     & 37  & 19    & 3800$\pm$2080 \\
Slit 2 & Middle 1-2 & 15.0$\pm$7.50     & 60  & 17    & 3500$\pm$1790 \\
Slit 2 & Middle 1-3 & 16.0$\pm$7.25     & 32  & 17    & 3900$\pm$1730 \\
Slit 2 & South-1    & 9.00$\pm$1.50     & 110 & 13    & 2200$\pm$364 \\
Slit 2 & South-2    & 7.50$\pm$2.00     & 46  & 13    & 1800$\pm$477 \\
Slit 2 & South-3    & 12.0$\pm$2.25     & 96  & 18    & 2900$\pm$524 \\
Slit 2 & South-4    & 14.0$\pm$2.25     & 34  & 14    & 3300$\pm$564 \\
Slit 2 & South-5    & 13.0$\pm$2.00     & 13  & 16    & 3000$\pm$455 \\
Slit 2 & South-6    & 1.30$\pm$4.75     & 14  & 12    & 300$\pm$1120 \\
Smith `91 & East-1  & 5.00$\pm$2.25     & 41  & 11    & 1200$\pm$540 \\
Smith `91 & East-2  & 2.80$\pm$1.75     & 36  & 11    & 660$\pm$435 \\
Smith `91 & East-3  & 10.0$\pm$3.50     & 15  & 10    & 2500$\pm$830 \\
Smith `91 & East-4  & 4.80$\pm$3.00     & 37  & 10    & 1100$\pm$725 \\
Smith `91 & East-5  & 11.0$\pm$1.75     & 14  & 13    & 2500$\pm$428 \\
Smith `91 & East-6  & 1.50$\pm$3.25     & 13  & 9     & 360$\pm$793 \\
Smith `91 & East-7  & 2.00$\pm$1.25     & 28  & 12    & 480$\pm$301 \\
Smith `91 & East-8  & 11.0$\pm$3.75     & 14  & 7     & 2700$\pm$885 \\
Smith `91 & East-9  & 5.50$\pm$2.50     & 9.7 & 9     & 1300$\pm$595 \\
Smith `91 & East-10 & 16.0$\pm$2.50     & 30  & 12    & 3900$\pm$567 \\
Smith `91 & East-11 & 0.50$\pm$1.75     & 31  & 8     & 120$\pm$401 \\
Smith `91 & East-12 & 3.50$\pm$1.25     & 15  & 9     & 840$\pm$310 \\
Smith `91 & East-13 & 5.00$\pm$2.00     & 20  & 9     & 1200$\pm$460 \\
Smith `91 & East-14 & 2.00$\pm$1.25     & 33  & 10    & 480$\pm$301 \\
Smith `91 & East-15 & 2.00$\pm$1.00     & 24  & 10    & 480$\pm$226 \\
Smith `91 & East-16 & 5.30$\pm$1.50     & 38  & 9     & 1300$\pm$342 \\
Smith `91 & East-17 & 11.0$\pm$6.00     & 12  & 9     & 2600$\pm$1420 \\
Smith `91 & East-18 & 7.30$\pm$3.50     & 18  & 12    & 1700$\pm$864 \\
    \enddata
        \vspace{-0.3cm} \tablecomments{$^1$ These velocities are the component in the plane of the sky.\\
        Best fit values of the shift and shock speed for each extraction aperture are followed by their $1\sigma$ statistical uncertainties.}
      \label{tab:expansions}  
    \end{deluxetable} 
\begin{figure*}
        \centering
        \includegraphics[scale=.55]{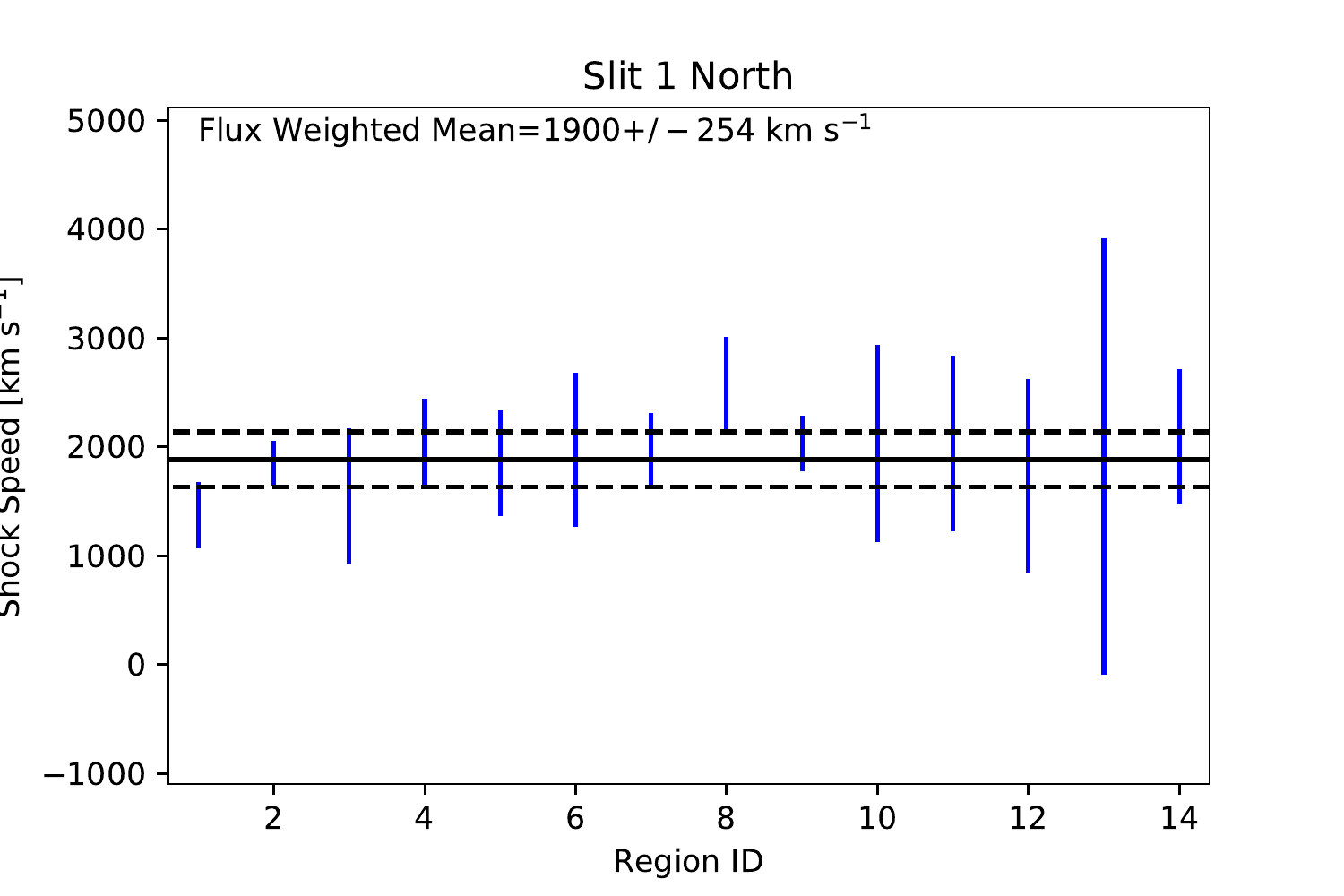}
        \includegraphics[scale=.55]{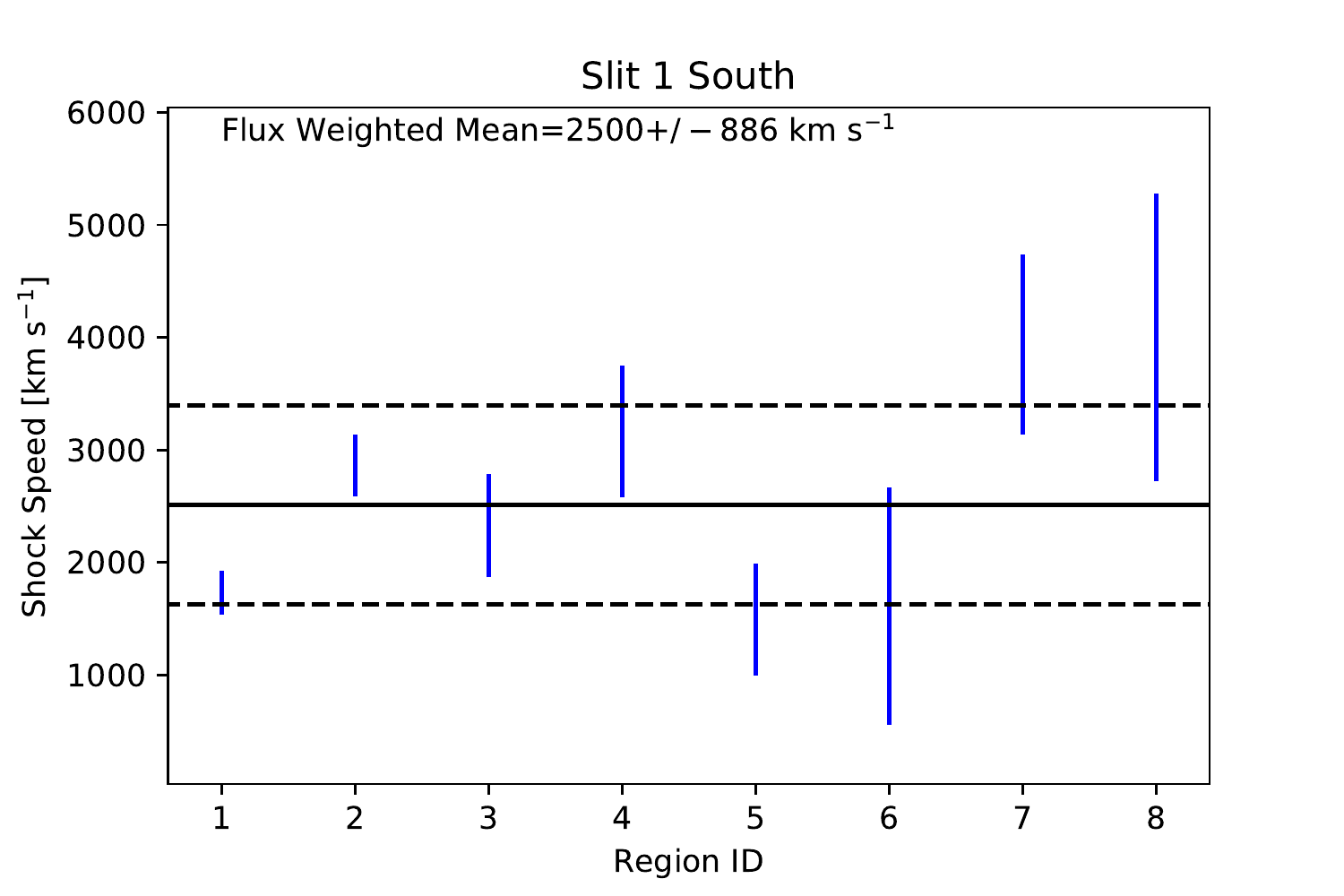}
        \includegraphics[scale=.55]{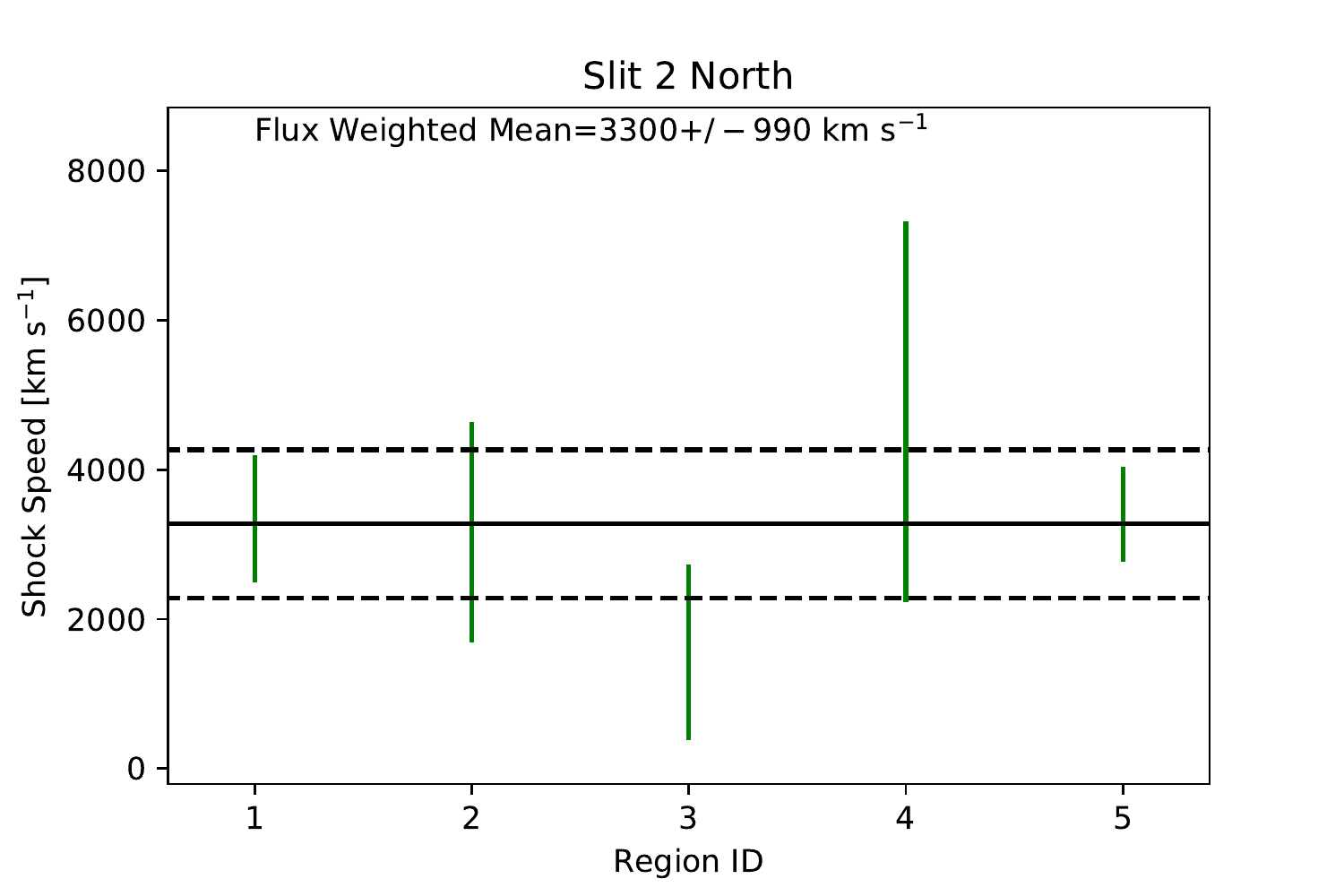}
        \includegraphics[scale=.55]{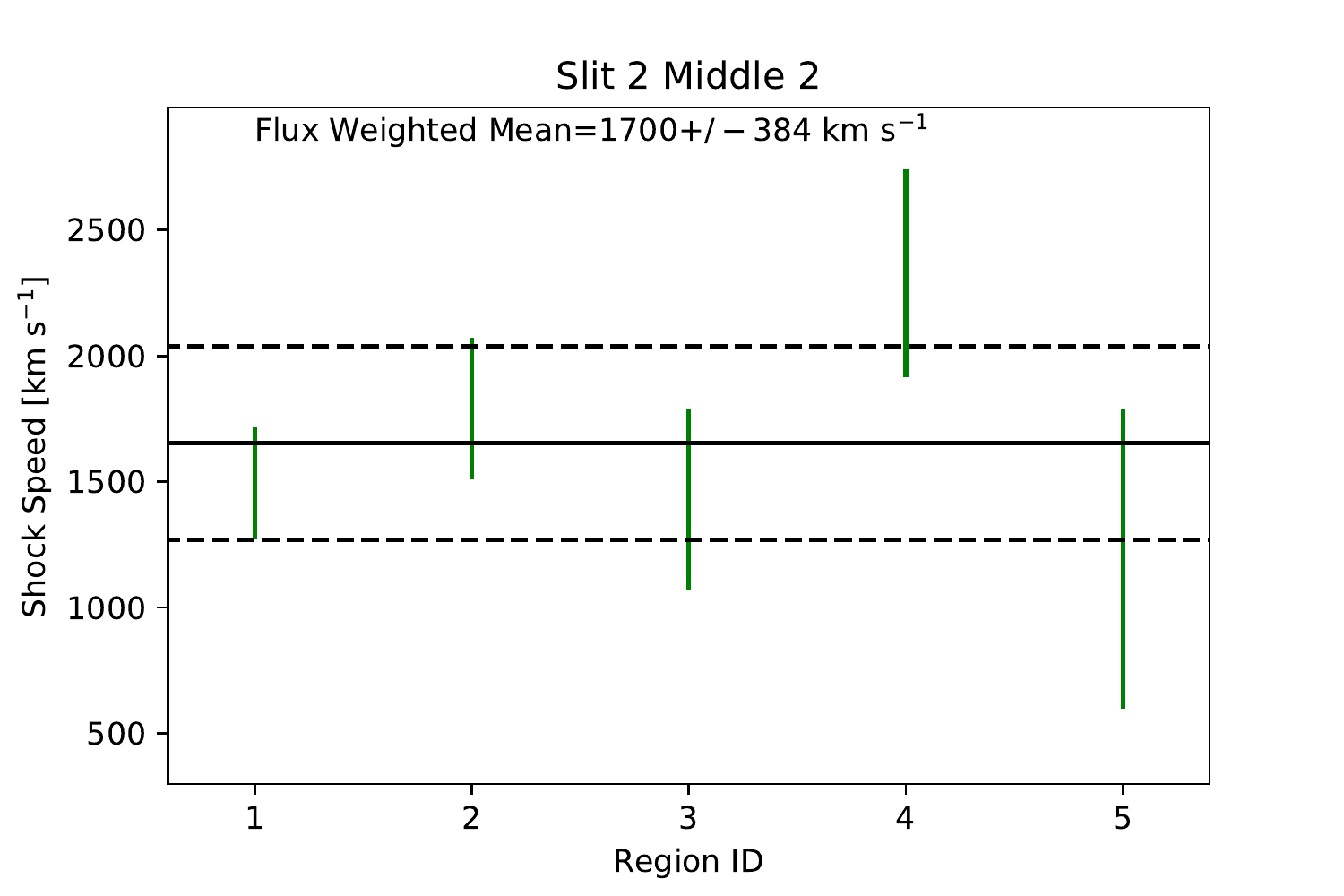}
        \includegraphics[scale=.55]{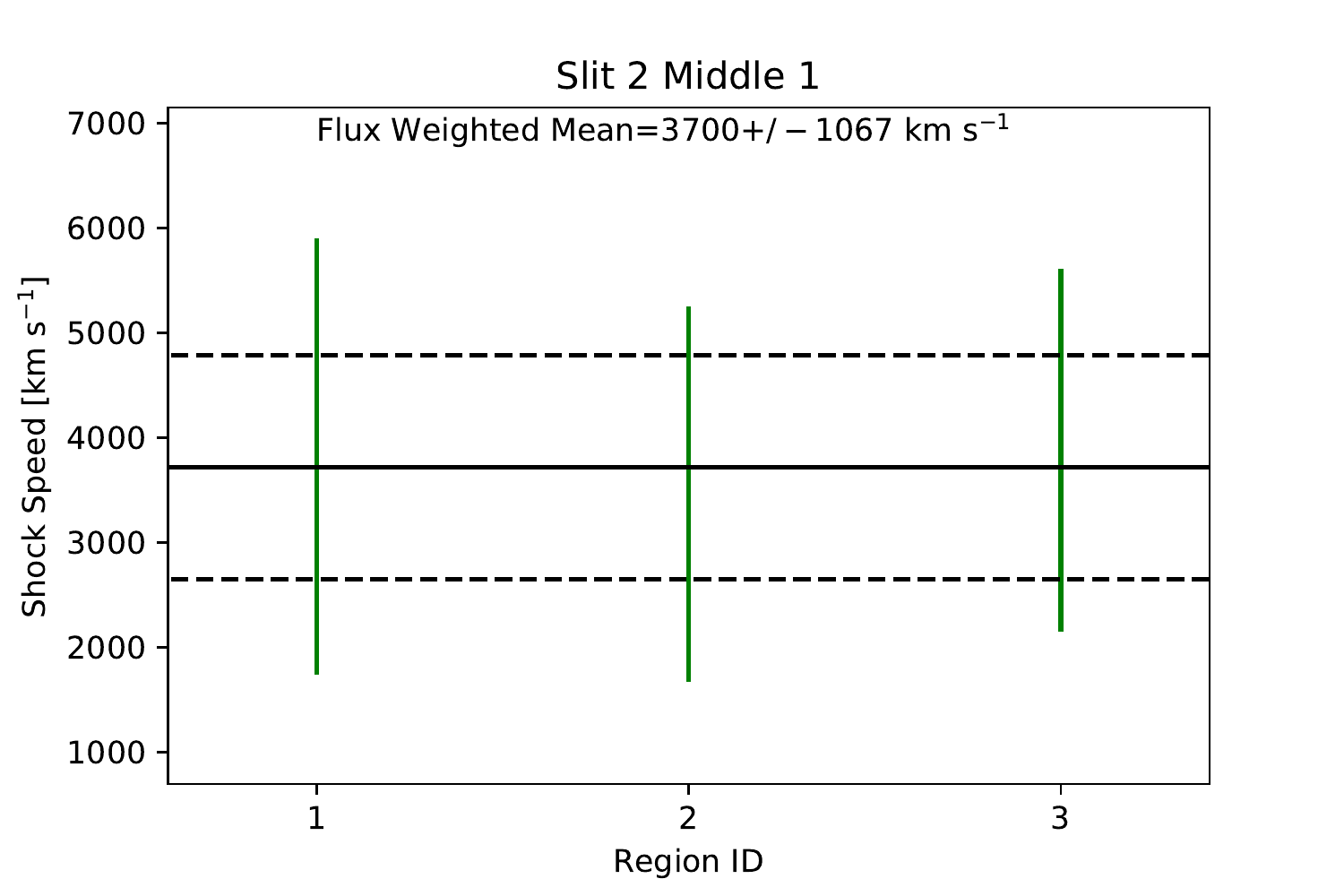}
        \includegraphics[scale=.55]{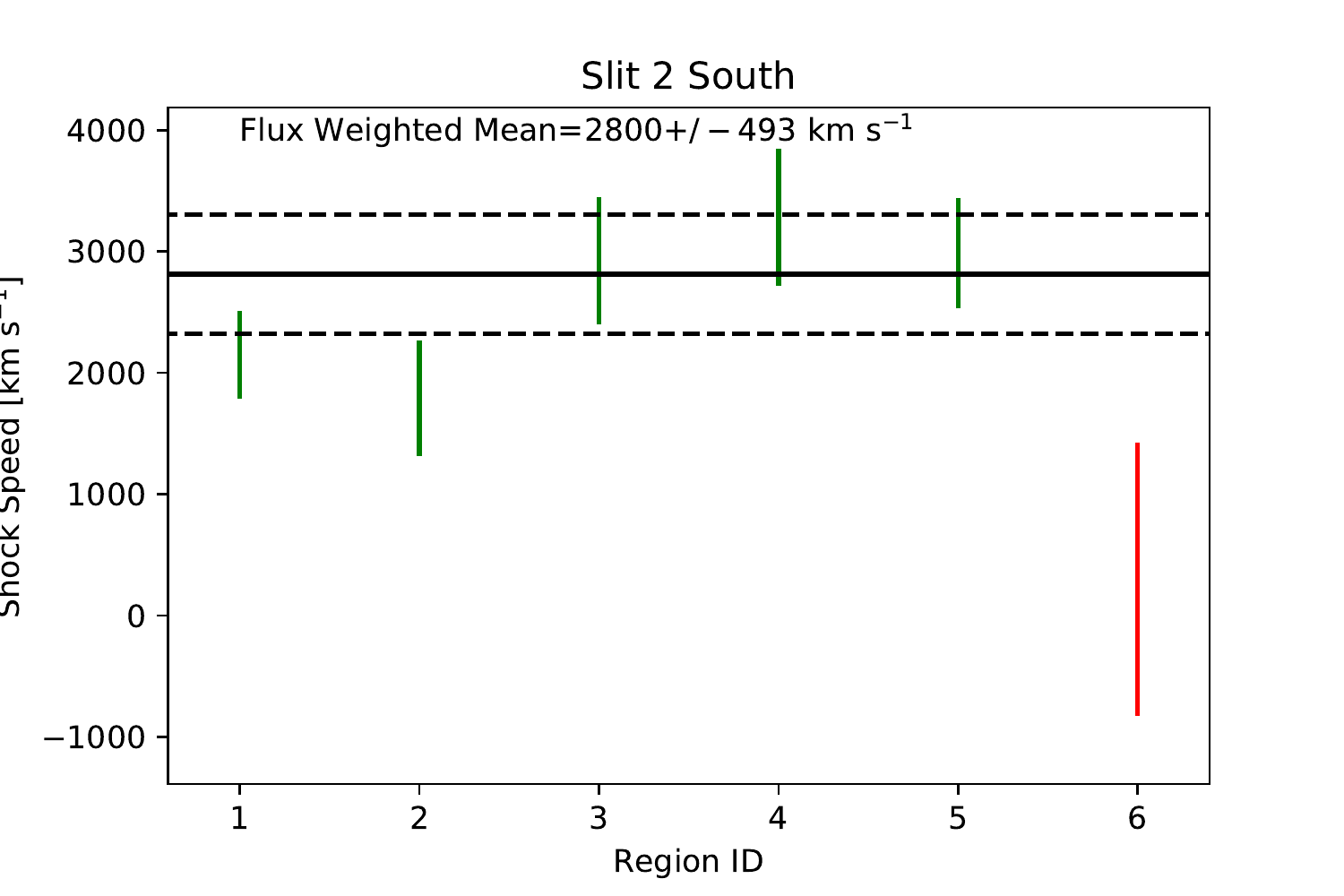}
        \includegraphics[scale=.55]{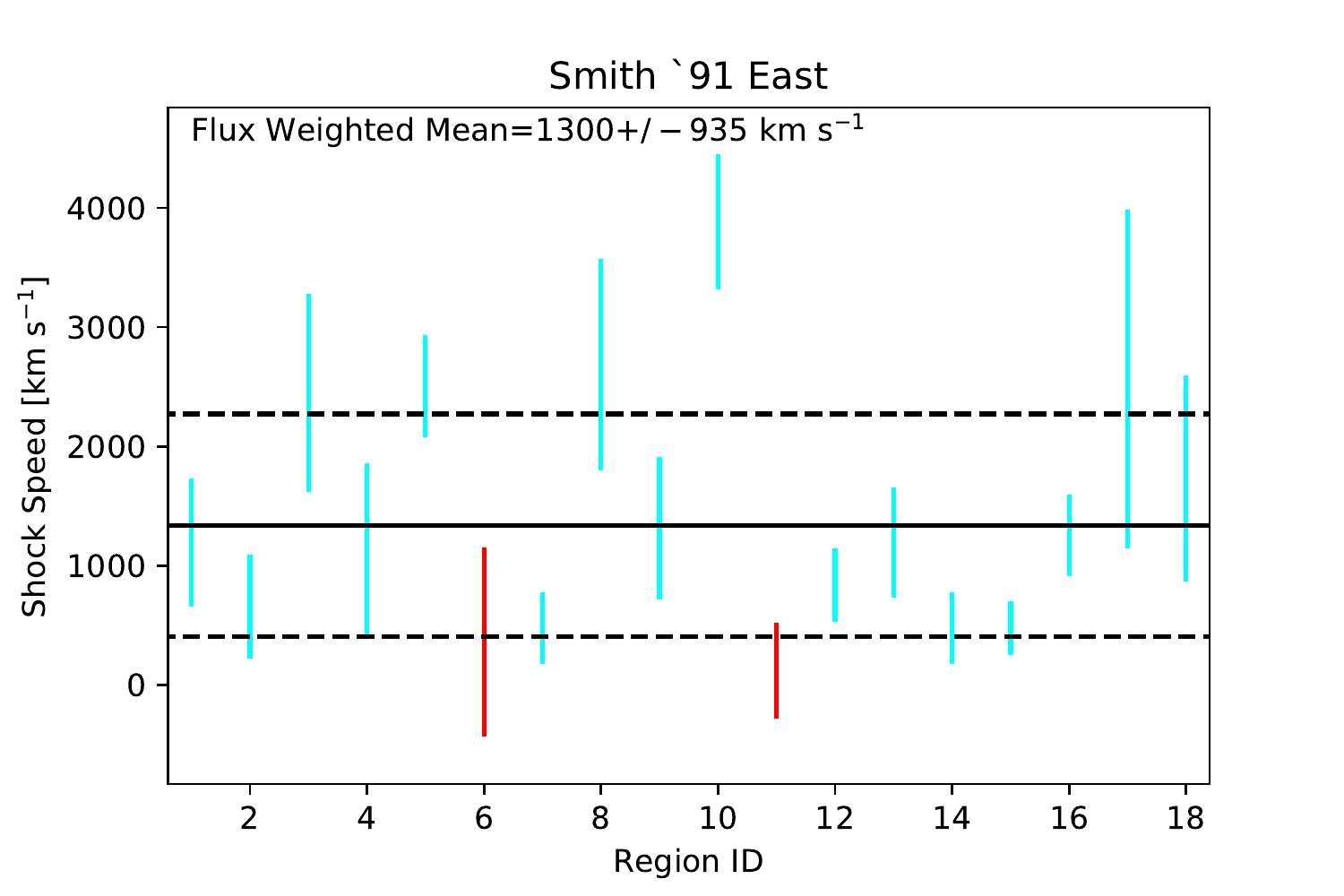}
        \caption{Shock speed determined with proper motion is plotted against respective region identification numbers for each extraction aperture in SNR \remtwo{}.  The solid black line shows the flux weighted mean with the 1$\sigma$ uncertainties shown with dashed black lines.  Data points in red have velocities below our spectral resolution at \ha{} of 400 \unit{}.}
         \label{fig:speeds.in.regions} 
    \end{figure*}
\clearpage

\section{Two-Dimensional Spectra of \remnant{} and \remtwo{}}
    \renewcommand\thefigure{A-\arabic{figure}}
        \begin{figure*}[ht]
            \centering
            \includegraphics[width=7.0truein]{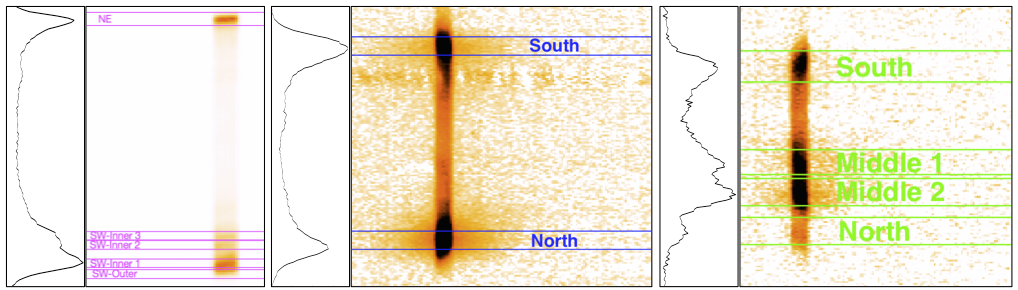}
            \caption{{\it left} - Two-dimensional FORS2 spectrum of the \ha{} emission line of \remnant{} with the extraction apertures labeled in magenta.  Hyperbolic-sin scaling is used to highlight inner filaments 2 and 3 in the southwestern portion of the remnant. {\it middle} - SALT RSS two dimensional spectrum of the \ha{} line extracted from slit 1 of \remtwo{}.  The north and south spectral extraction aperture is shown in blue.  {\it right} -  Slit 2 RSS two-dimensional spectrum of the \ha{}  line for the south middle 1, middle 2, and north spectral extraction apertures is shown in green.}
            \label{fig:2d_spec}
        \end{figure*}  
    
    The spectral extraction apertures in \remnant{} are 1.0$^{\prime\prime}$ wide in the spatial direction in the southwest, and 1.5$^{\prime\prime}$ in the northeast.  We use the star at the WCS position \traceone{}, and will reference the centers of the spectral extraction regions as distances from said star.  In the southwest the extraction regions for the outer filament and inner filaments 1-3 are 4.2$^{\prime\prime}$, 5.5$^{\prime\prime}$, 7.7$^{\prime\prime}$, and 8.7$^{\prime\prime}$ away from the trace star respectively, while the northeast extraction region is at a distance of 33.7$^{\prime\prime}$.    
    
    In the case of slit 1 for \remtwo{} we use a trace star, which was our guide star for the observation, at the position \tracetwo{}, and the slit is oriented at an angle of 55$^{\circ}$ west of north.  The northern and southern extraction apertures are at distance of 9.9$^{\prime\prime}$ and 36.4$^{\prime\prime}$ respectively from the aforementioned trace star.  The trace/guide star for slit 2 is at a position of \tracethree{} and the aperture sizes (in the spatial direction) for the North, middle 2, middle 1, and south regions are 2.5$^{\prime\prime}$, 3.5$^{\prime\prime}$, 3.3$^{\prime\prime}$, and 4$^{\prime\prime}$ respectively.  Slit 2 is oriented at an angle of 60$^{\circ}$ west of north.  The centers of these apertures are at distances of 12.8$^{\prime\prime}$, 18.3$^{\prime\prime}$, 22.6$^{\prime\prime}$, and 34.3$^{\prime\prime}$ from north to south respectively for the four regions.

     \section{Spectral fits of \ha{} line for \remnant{} and \remtwo{}}
     
    \begin{figure*}[ht]
        \centering
        \includegraphics[scale=.38]{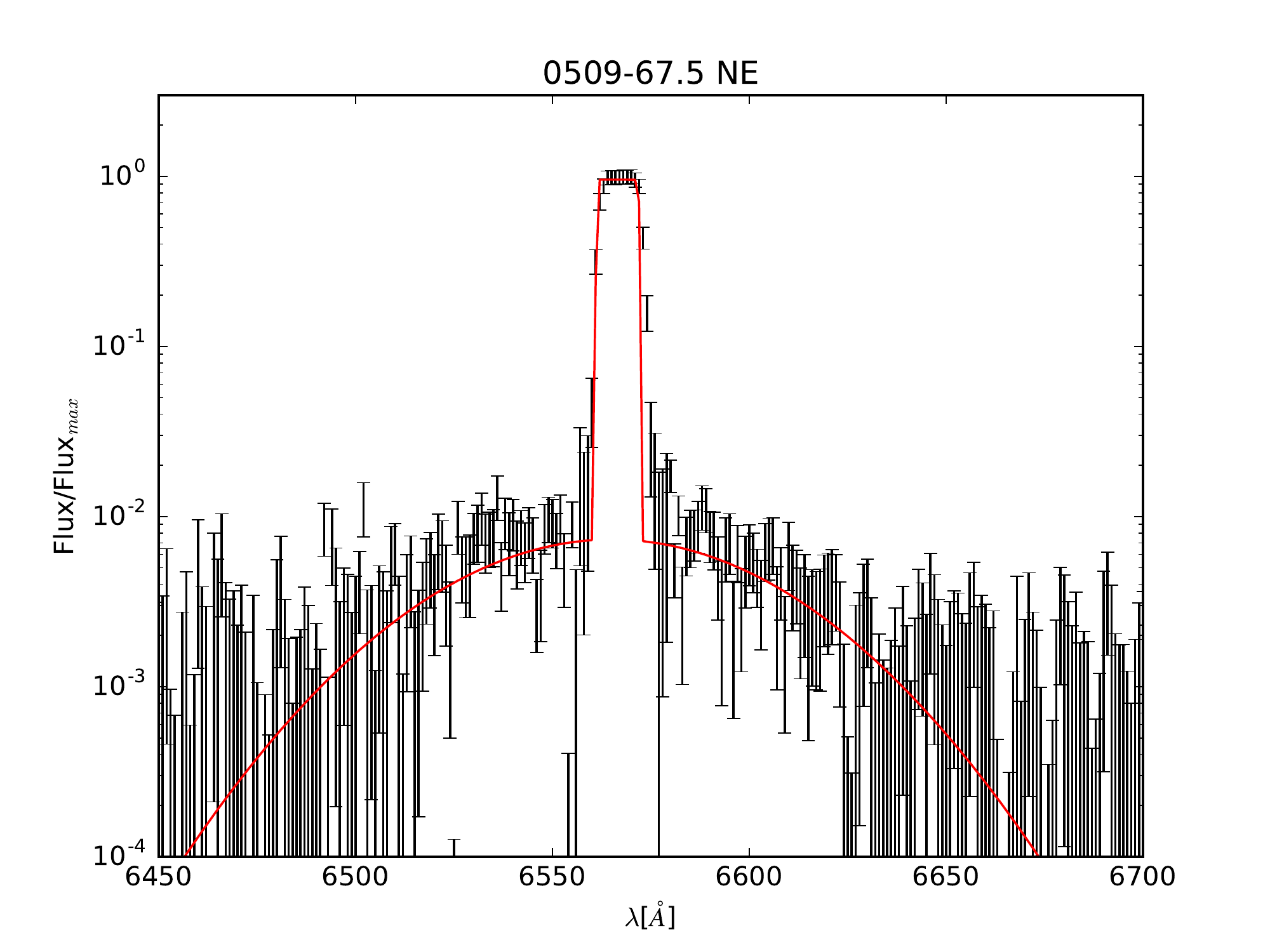}
        \includegraphics[scale=.38]{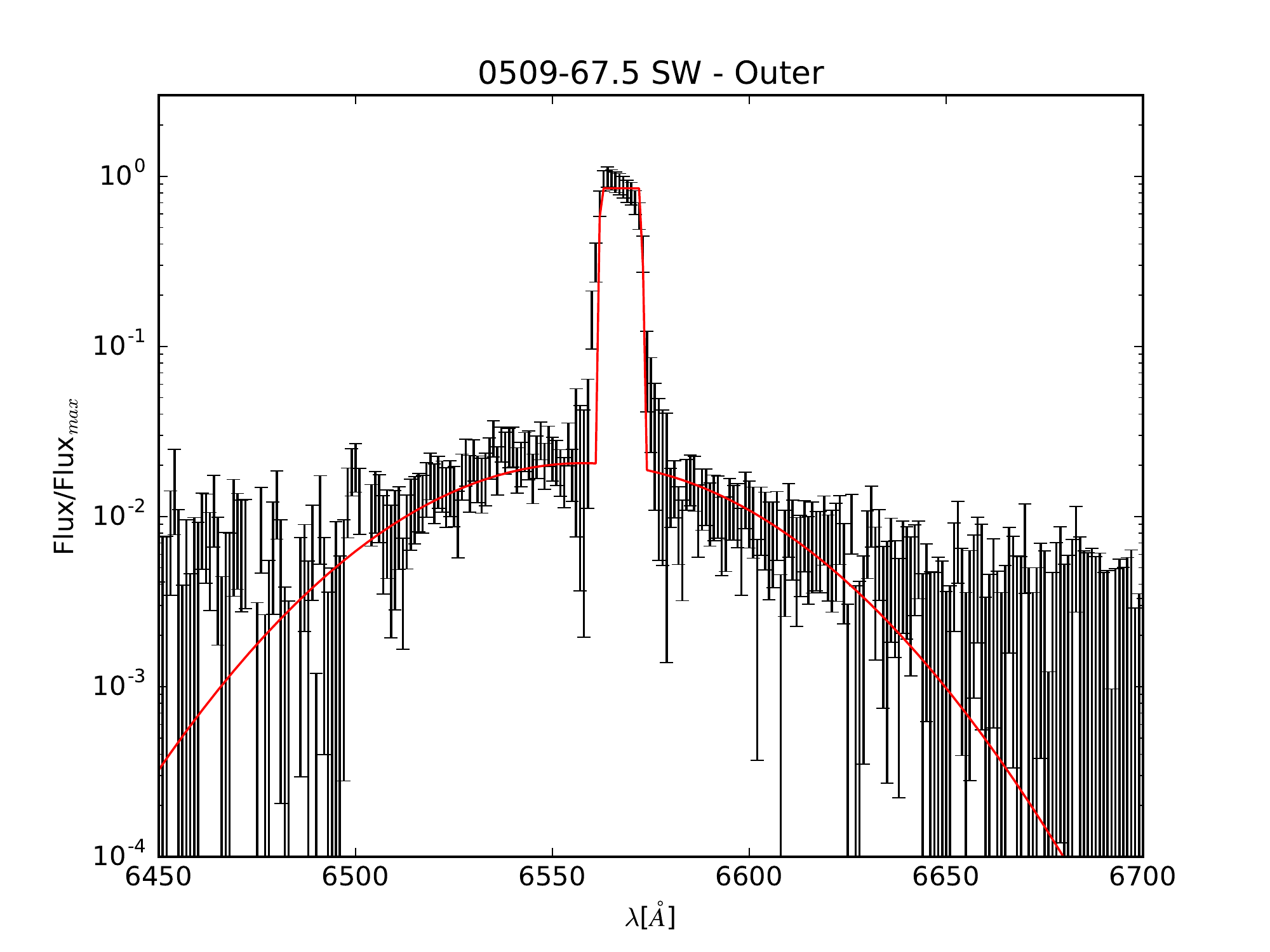}
    \end{figure*}
        
    \begin{figure*}[h]
        \centering
        \includegraphics[scale=.38]{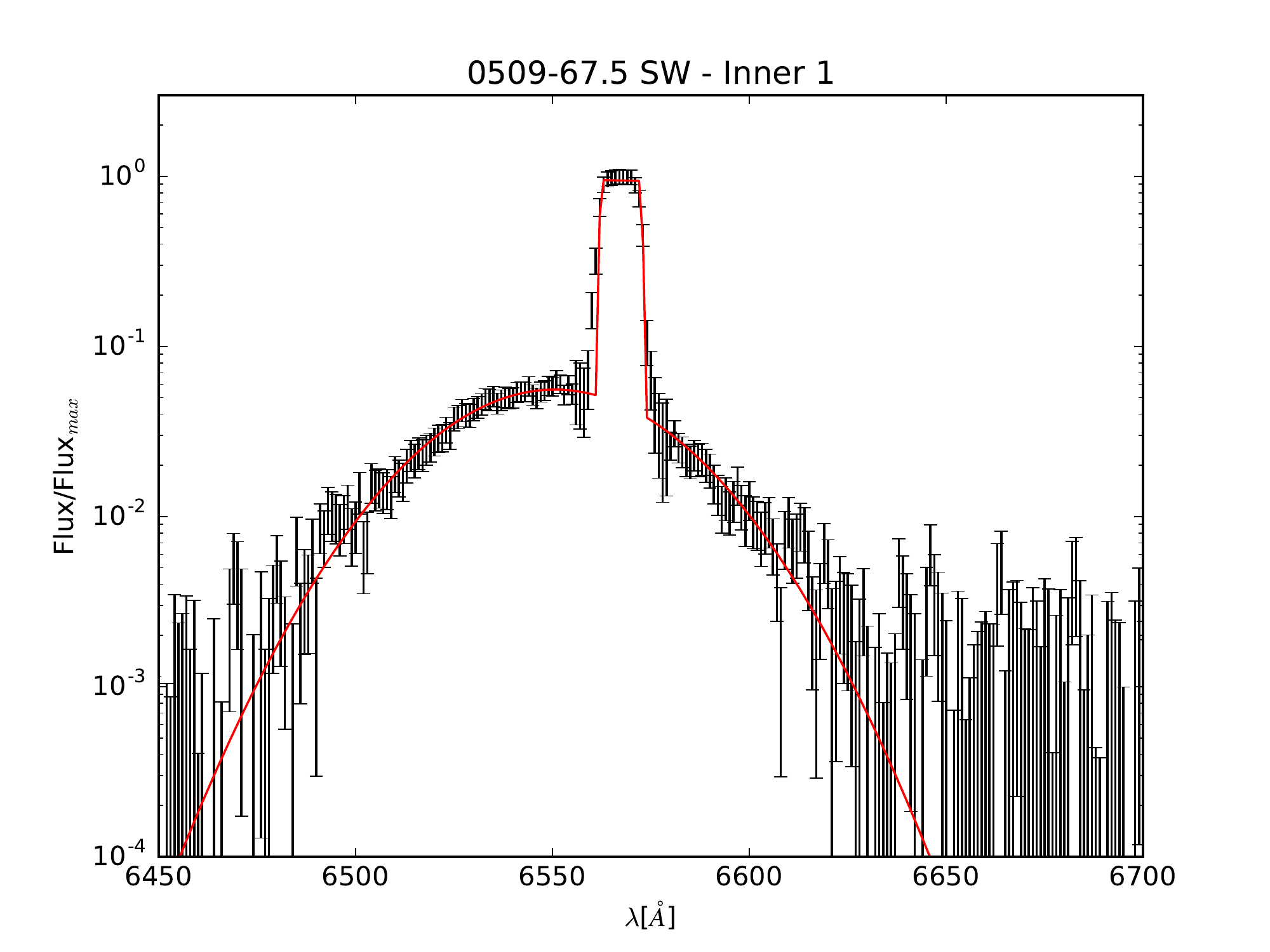}
        \includegraphics[scale=.38]{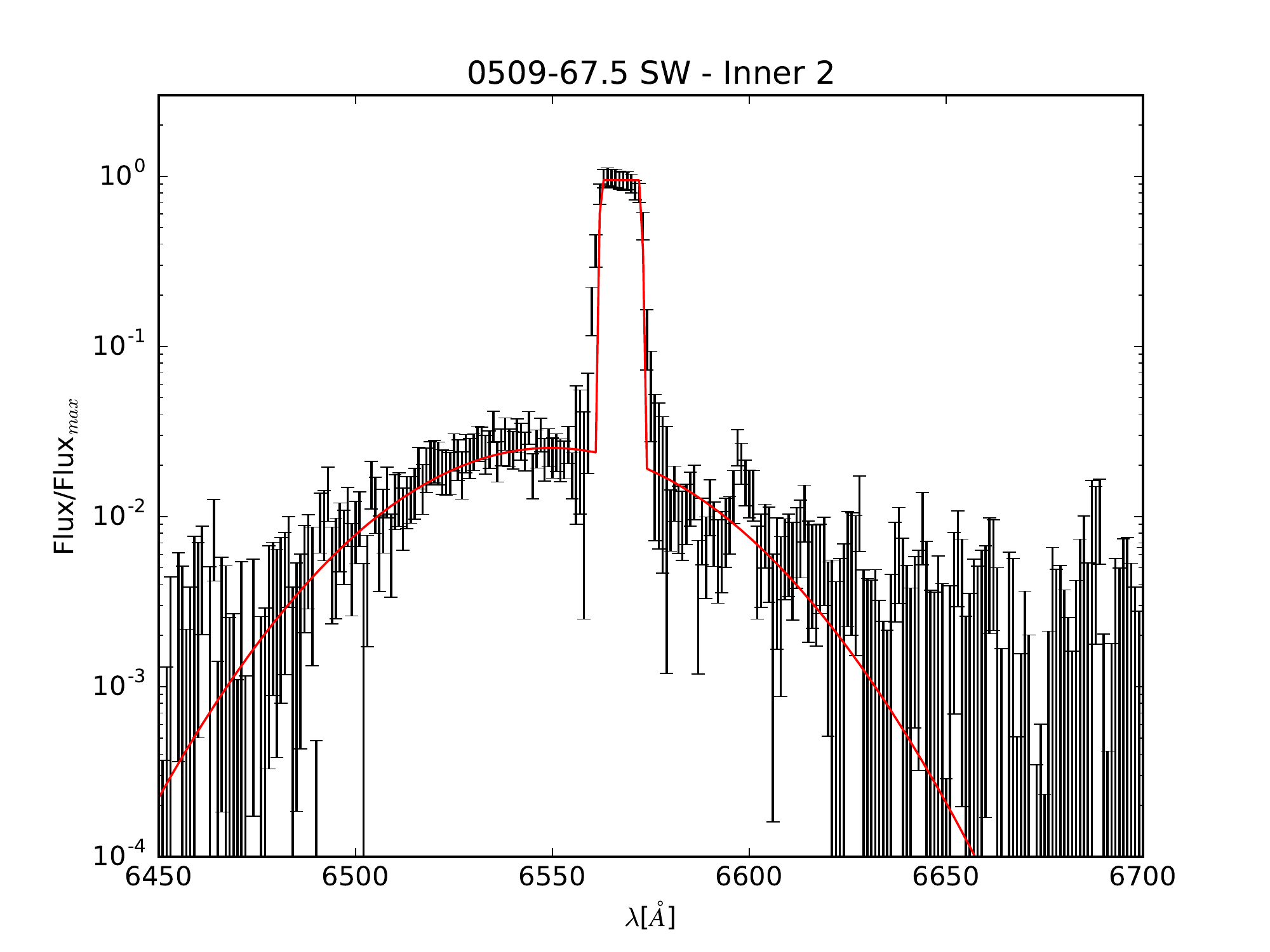}
        \includegraphics[scale=.38]{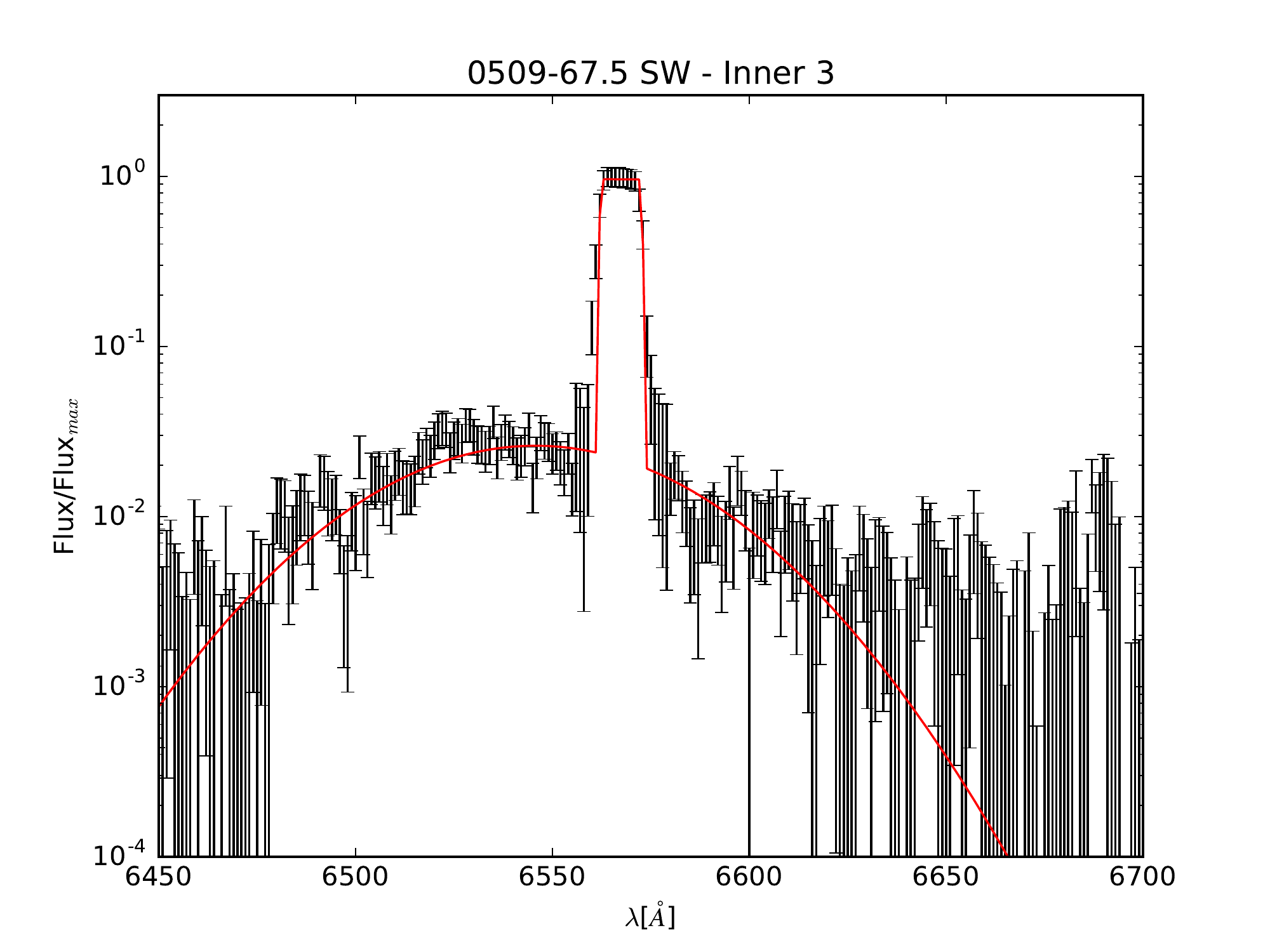}
    \caption{Spectra of the \ha{} line for \remnant{} with the FORS2 spectrograph.  Our best fit spectrum is shown for each panel with a red curve. }
        \label{fig:0509_spectra}  
    \end{figure*}
    
    \begin{figure*}[h]
        \centering
        \includegraphics[scale=.38]{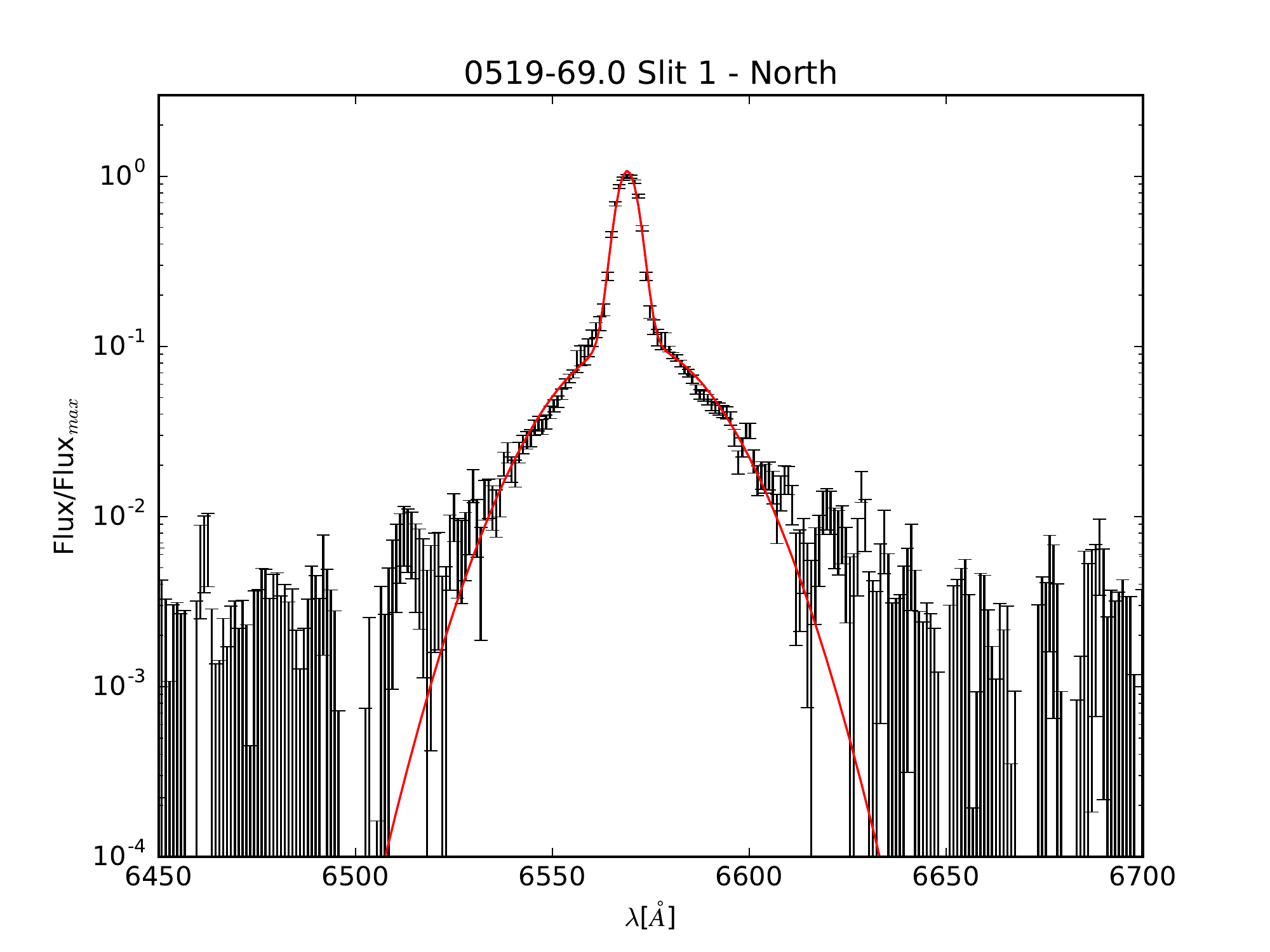}
        \includegraphics[scale=.38]{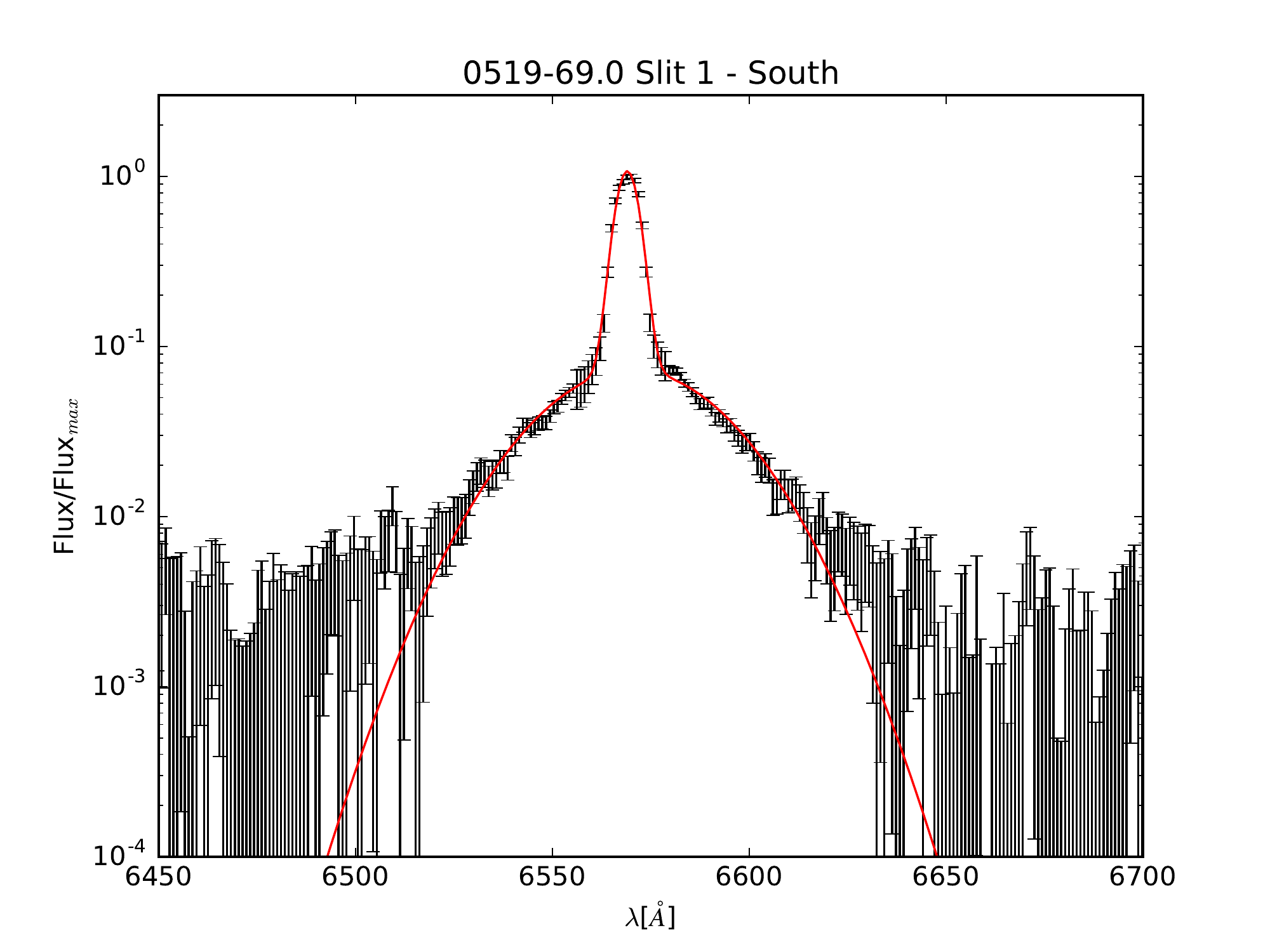}
    \end{figure*}
         
    \begin{figure*}
        \centering
        \includegraphics[scale=.38]{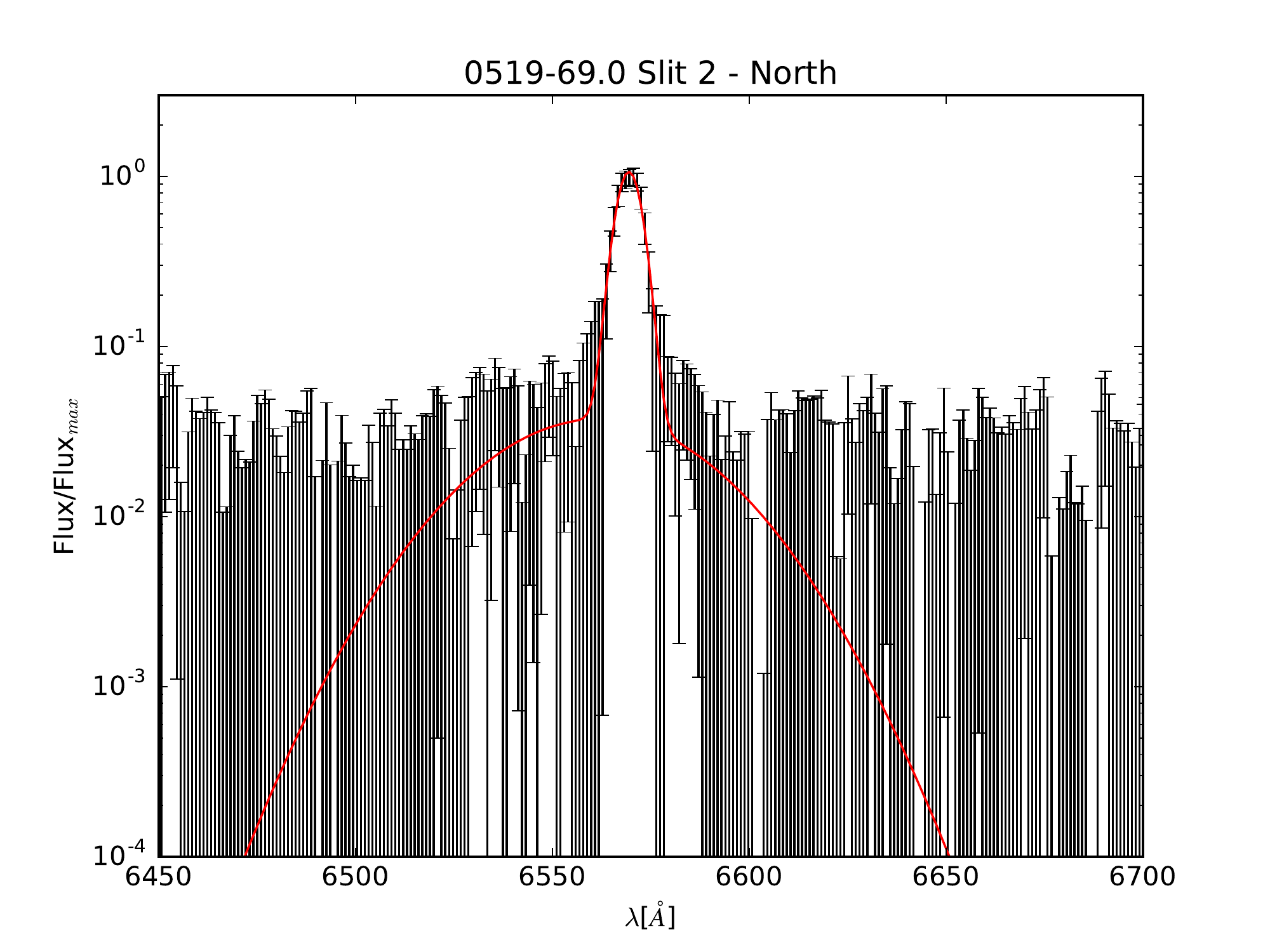}
        \includegraphics[scale=.38]{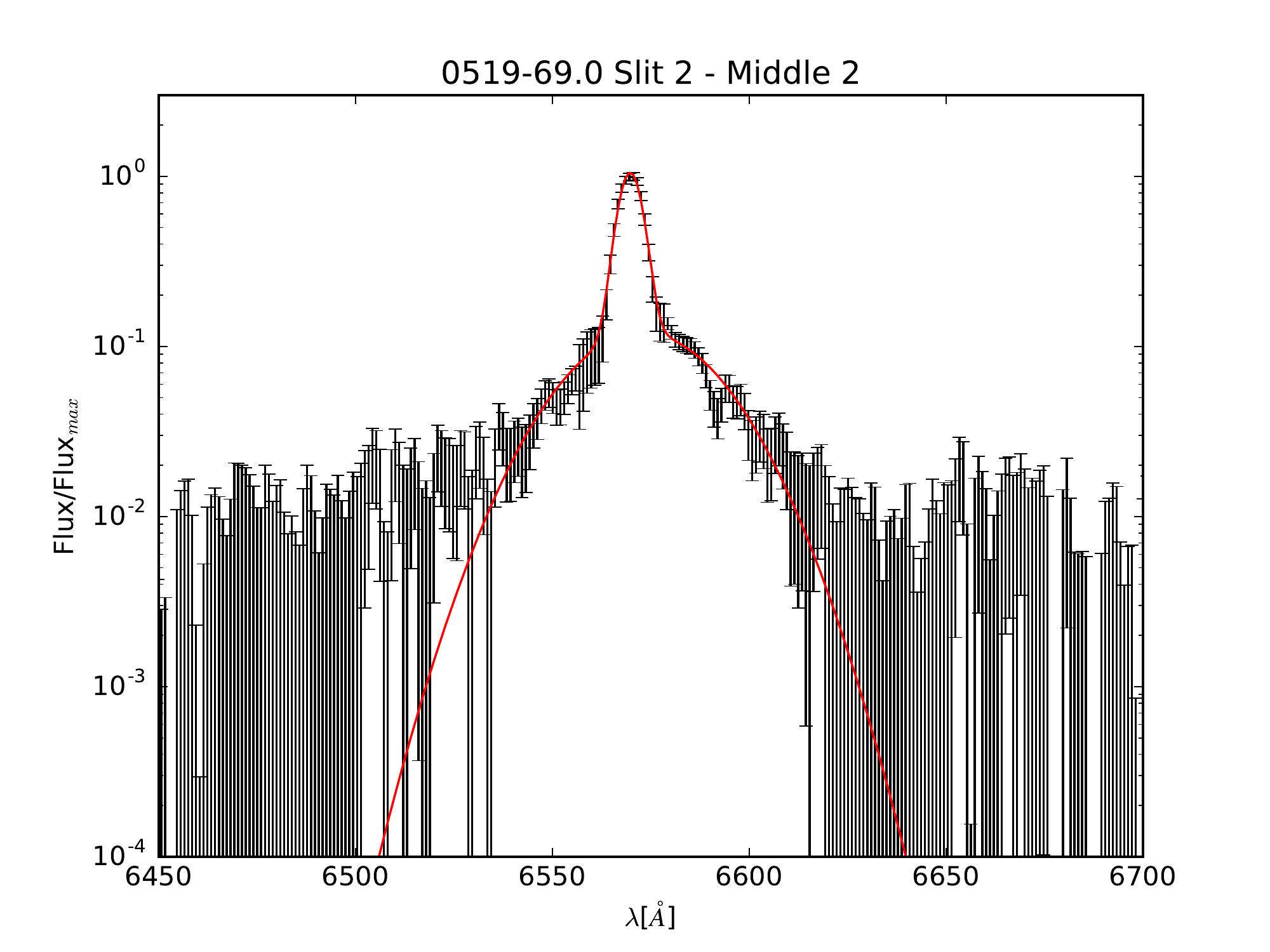}
        \includegraphics[scale=.38]{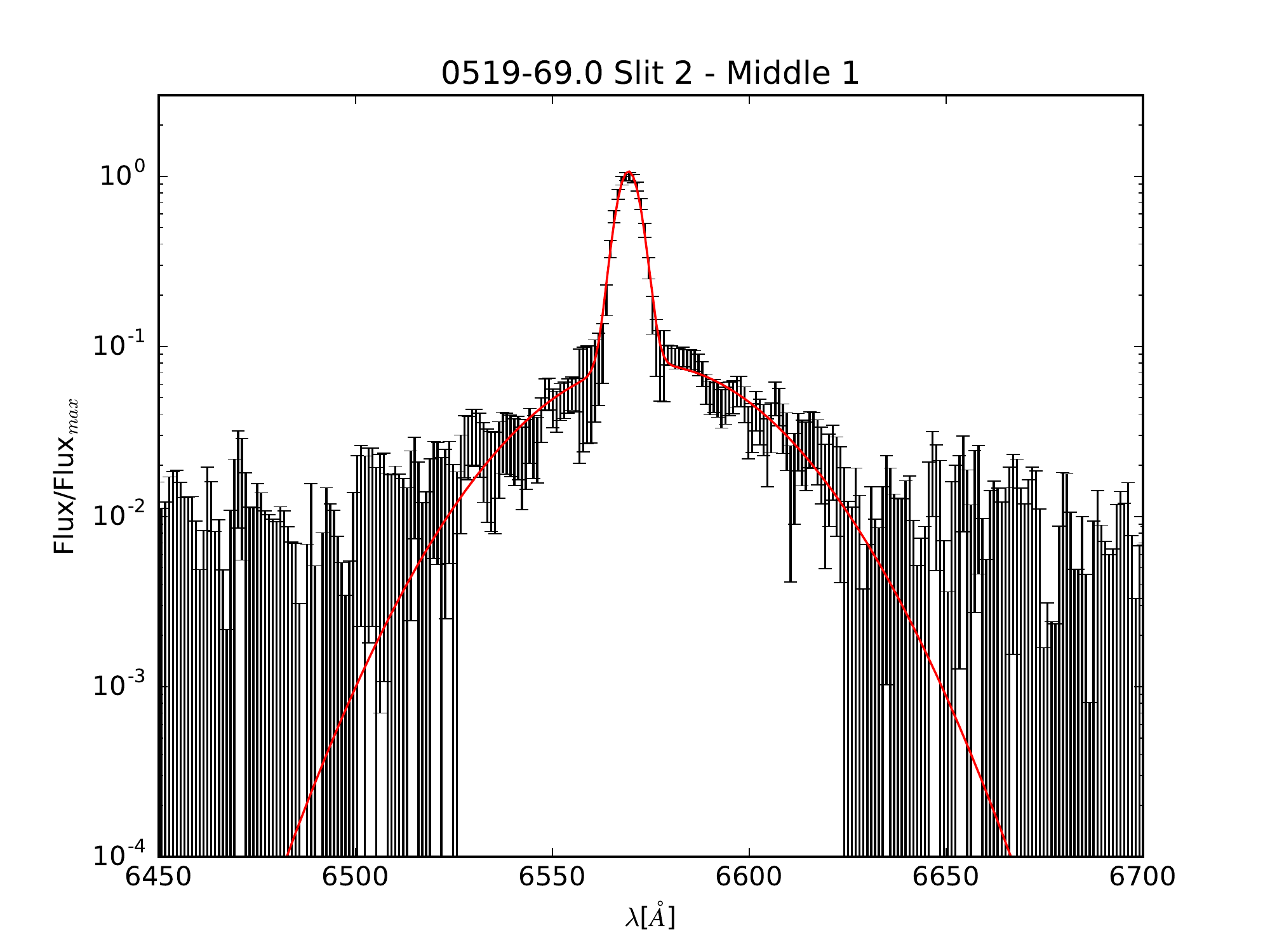}
        \includegraphics[scale=.38]{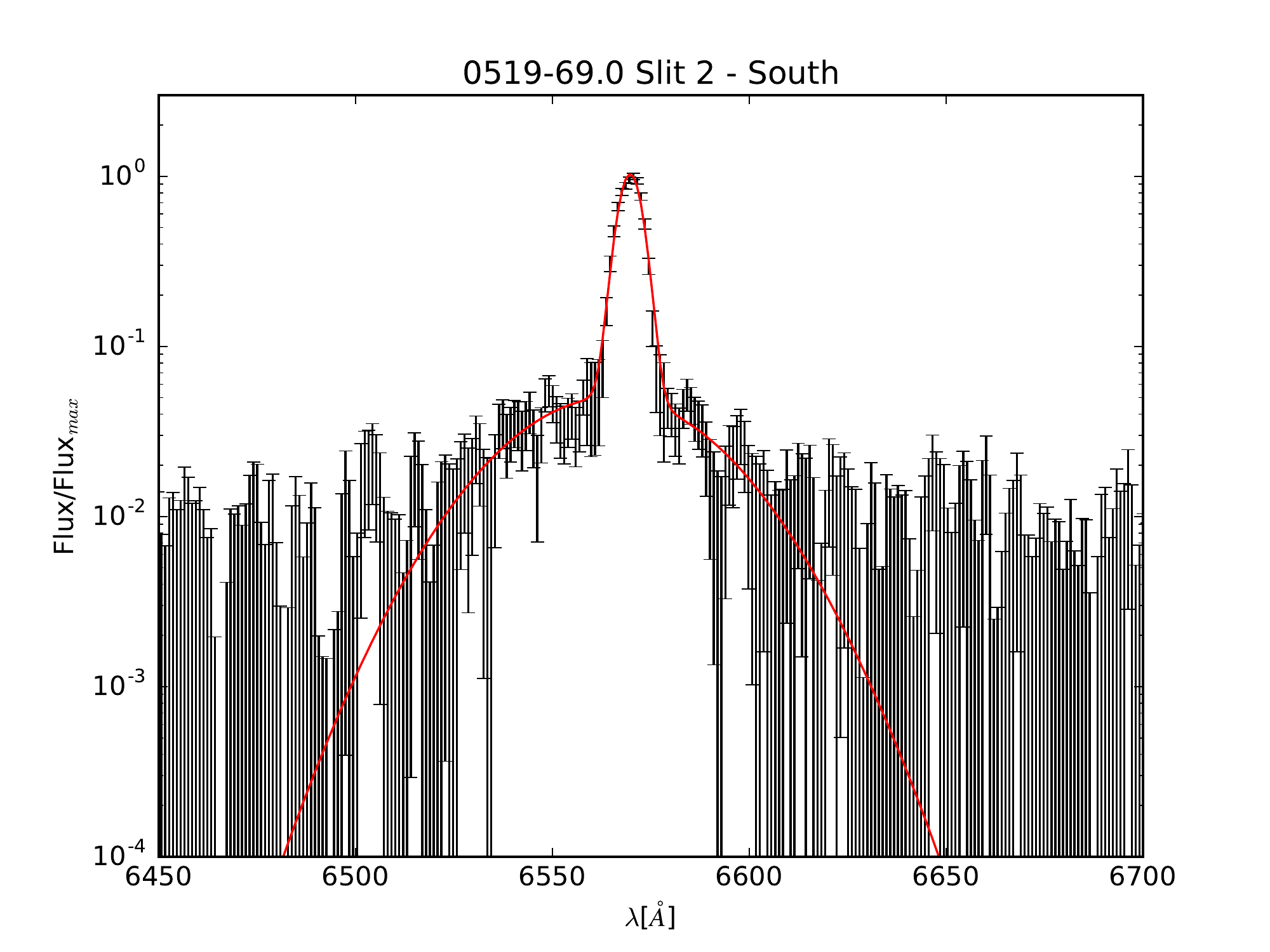}
        \caption{SALT RSS spectra of the \ha{} line from the two longslit positions of \remtwo{}.  Our best fit spectrum is shown for each panel with a red curve. }
        \label{fig:0519_spectra}
    \end{figure*}


\begin{thebibliography}{}
\expandafter\ifx\csname natexlab\endcsname\relax\def\natexlab#1{#1}\fi

\bibitem[Kne{\v z}evi{\'c} et al.(2017)]{knezevic2017} Kne{\v z}evi{\'c}, S., L{\"a}sker, R., van de Ven, G., et al.\ 2017, \apj, 846, 167 

\bibitem[Morlino et al.(2013)]{morlino2013} Morlino, G., Blasi, P., Bandiera, R., Amato, E., \& Caprioli, D.\ 2013, \apj, 768, 148 

\bibitem[Marcowith et al.(2016)]{marcowith2016} Marcowith, A., Bret, A., Bykov, A., et al.\ 2016, Reports on Progress in Physics, 79, 046901 

\bibitem[{{Abdo} {et~al.}(2010){Abdo}, {Ackermann}, {Ajello}, {Allafort},
  {Baldini}, {Ballet}, {Barbiellini}, {Baring}, {Bastieri}, {Baughman},
  {Bechtol}, {Bellazzini}, {Berenji}, {Blandford}, {Bloom}, {Bonamente},
  {Borgland}, {Bregeon}, {Brez}, {Brigida}, {Bruel}, {Buehler}, {Burnett},
  {Busetto}, {Caliandro}, {Cameron}, {Caraveo}, {Casandjian}, {Cecchi}, {{\c
  C}elik}, {Charles}, {Chaty}, {Chekhtman}, {Cheung}, {Chiang}, {Cillis},
  {Ciprini}, {Claus}, {Cohen-Tanugi}, {Conrad}, {Corbel}, {de Palma}, {Digel},
  {Dormody}, {Silva}, {Drell}, {Dubois}, {Dumora}, {Edmonds}, {Farnier},
  {Favuzzi}, {Fegan}, {Ferrara}, {Focke}, {Fortin}, {Frailis}, {Fukazawa},
  {Funk}, {Fusco}, {Gargano}, {Gasparrini}, {Gehrels}, {Germani}, {Giavitto},
  {Giglietto}, {Giordano}, {Glanzman}, {Godfrey}, {Grenier}, {Grondin},
  {Grove}, {Guillemot}, {Guiriec}, {Hanabata}, {Hays}, {Harding}, {Hayashida},
  {Horan}, {Hughes}, {Jackson}, {Johnson}, {Johnson}, {Johnson}, {Kamae},
  {Katagiri}, {Kataoka}, {Kawai}, {Kerr}, {Kn{\"o}dlseder}, {Kuss}, {Lande},
  {Latronico}, {Lemoine-Goumard}, {Longo}, {Loparco}, {Lott}, {Lovellette},
  {Lubrano}, {Makeev}, {Mazziotta}, {Meurer}, {Michelson}, {Mitthumsiri},
  {Mizuno}, {Monte}, {Monzani}, {Morselli}, {Moskalenko}, {Murgia}, {Nakamori},
  {Nolan}, {Norris}, {Nuss}, {Ohsugi}, {Okumura}, {Omodei}, {Orlando}, {Ormes},
  {Paneque}, {Panetta}, {Pelassa}, {Pepe}, {Pesce-Rollins}, {Piron}, {Pohl},
  {Porter}, {Rain{\`o}}, {Rando}, {Reimer}, {Reimer}, {Reposeur}, {Ritz},
  {Rodriguez}, {Romani}, {Roth}, {Sadrozinski}, {Sander}, {Saz Parkinson},
  {Scargle}, {Sgr{\`o}}, {Siskind}, {Smith}, {Smith}, {Spinelli}, {Strickman},
  {Suson}, {Tajima}, {Takahashi}, {Tanaka}, {Thayer}, {Thayer}, {Thompson},
  {Thorsett}, {Tibaldo}, {Tibolla}, {Torres}, {Tosti}, {Tramacere}, {Uchiyama},
  {Usher}, {Van Etten}, {Vasileiou}, {Venter}, {Vilchez}, {Vitale}, {Waite},
  {Wang}, {Winer}, {Wood}, {Yamazaki}, {Ylinen}, \& {Ziegler}}]{abdo2010}
{Abdo}, A.~A., {Ackermann}, M., {Ajello}, M., {et~al.} 2010, \apjl, 710, L92

\bibitem[{{Acero} {et~al.}(2010){Acero}, {Aharonian}, {Akhperjanian}, {Anton},
  {Barres de Almeida}, {Bazer-Bachi}, {Becherini}, {Behera}, {Beilicke},
  {Bernl{\"o}hr}, {Bochow}, {Boisson}, {Bolmont}, {Borrel}, {Brucker}, {Brun},
  {Brun}, {B{\"u}hler}, {Bulik}, {B{\"u}sching}, {Boutelier}, {Chadwick},
  {Charbonnier}, {Chaves}, {Cheesebrough}, {Conrad}, {Chounet}, {Clapson},
  {Coignet}, {Dalton}, {Daniel}, {Davids}, {Degrange}, {Deil}, {Dickinson},
  {Djannati-Ata{\"\i}}, {Domainko}, {O'C.~Drury}, {Dubois}, {Dubus}, {Dyks},
  {Dyrda}, {Egberts}, {Eger}, {Espigat}, {Fallon}, {Farnier}, {Fegan},
  {Feinstein}, {Fiasson}, {F{\"o}rster}, {Fontaine}, {F{\"u}{\ss}ling},
  {Gabici}, {Gallant}, {G{\'e}rard}, {Gerbig}, {Giebels}, {Glicenstein},
  {Gl{\"u}ck}, {Goret}, {G{\"o}ring}, {Hauser}, {Hauser}, {Heinz},
  {Heinzelmann}, {Henri}, {Hermann}, {Hinton}, {Hoffmann}, {Hofmann},
  {Hofverberg}, {Holleran}, {Hoppe}, {Horns}, {Jacholkowska}, {de Jager},
  {Jahn}, {Jung}, {Katarzy{\'n}ski}, {Katz}, {Kaufmann}, {Kerschhaggl},
  {Khangulyan}, {Kh{\'e}lifi}, {Keogh}, {Klochkov}, {Klu{\'z}niak}, {Kneiske},
  {Komin}, {Kosack}, {Kossakowski}, {Lamanna}, {Lemoine-Goumard}, {Lenain},
  {Lohse}, {Marandon}, {Marcowith}, {Masbou}, {Maurin}, {McComb}, {Medina},
  {M{\'e}hault}, {Moderski}, {Moulin}, {Naumann-Godo}, {de Naurois}, {Nedbal},
  {Nekrassov}, {Nicholas}, {Niemiec}, {Nolan}, {Ohm}, {Olive}, {de O{\~n}a
  Wilhelmi}, {Orford}, {Ostrowski}, {Panter}, {Paz Arribas}, {Pedaletti},
  {Pelletier}, {Petrucci}, {Pita}, {P{\"u}hlhofer}, {Punch}, {Quirrenbach},
  {Raubenheimer}, {Raue}, {Rayner}, {Reimer}, {Renaud}, {de Los Reyes},
  {Rieger}, {Ripken}, {Rob}, {Rosier-Lees}, {Rowell}, {Rudak}, {Rulten},
  {Ruppel}, {Ryde}, {Sahakian}, {Santangelo}, {Schlickeiser}, {Sch{\"o}ck},
  {Sch{\"o}nwald}, {Schwanke}, {Schwarzburg}, {Schwemmer}, {Shalchi}, {Sushch},
  {Sikora}, {Skilton}, {Sol}, {Stawarz}, {Steenkamp}, {Stegmann}, {Stinzing},
  {Superina}, {Szostek}, {Tam}, {Tavernet}, {Terrier}, {Tibolla}, {Tluczykont},
  {van Eldik}, {Vasileiadis}, {Venter}, {Venter}, {Vialle}, {Vincent}, {Vink},
  {Vivier}, {V{\"o}lk}, {Volpe}, {Vorobiov}, {Wagner}, {Ward}, {Zdziarski},
  {Zech}, \& {H.E.S.S.~Collaboration}}]{acero2010}
{Acero}, F., {Aharonian}, F., {Akhperjanian}, A.~G., {et~al.} 2010, \aap, 516,
  A62

\bibitem[{{Aharonian} {et~al.}(2009){Aharonian}, {Akhperjanian}, {de Almeida},
  {Bazer-Bachi}, {Behera}, {Beilicke}, {Benbow}, {Bernl{\"o}hr}, {Boisson},
  {Bochow}, {Borrel}, {Braun}, {Brion}, {Brucker}, {B{\"u}hler}, {Bulik},
  {B{\"u}sching}, {Boutelier}, {Carrigan}, {Chadwick}, {Charbonnier}, {Chaves},
  {Chounet}, {Clapson}, {Coignet}, {Costamante}, {Dalton}, {Degrange},
  {Dickinson}, {Djannati-Ata{\"\i}}, {Domainko}, {Drury}, {Dubois}, {Dubus},
  {Dyks}, {Egberts}, {Emmanoulopoulos}, {Espigat}, {Farnier}, {Feinstein},
  {Fiasson}, {F{\"o}rster}, {Fontaine}, {F{\"u}{\ss}ling}, {Gabici}, {Gallant},
  {G{\'e}rard}, {Giebels}, {Glicenstein}, {Gl{\"u}ck}, {Goret},
  {Hadjichristidis}, {Hauser}, {Hauser}, {Heinzelmann}, {Henri}, {Hermann},
  {Hinton}, {Hoffmann}, {Hofmann}, {Holleran}, {Hoppe}, {Horns},
  {Jacholkowska}, {de Jager}, {Jung}, {Katarzy{\'n}ski}, {Kaufmann},
  {Kendziorra}, {Kerschhaggl}, {Khangulyan}, {Kh{\'e}lifi}, {Keogh}, {Komin},
  {Kosack}, {Lamanna}, {Latham}, {Lemoine-Goumard}, {Lenain}, {Lohse},
  {Marandon}, {Martin}, {Martineau-Huynh}, {Marcowith}, {Masterson}, {Maurin},
  {McComb}, {Medina}, {Moderski}, {Moulin}, {Naumann-Godo}, {de Naurois},
  {Nedbal}, {Nekrassov}, {Niemiec}, {Nolan}, {Ohm}, {Olive}, {de O{\~n}a
  Wilhelmi}, {Orford}, {Osborne}, {Ostrowski}, {Panter}, {Pedaletti},
  {Pelletier}, {Petrucci}, {Pita}, {P{\"u}hlhofer}, {Punch}, {Quirrenbach},
  {Raubenheimer}, {Raue}, {Rayner}, {Renaud}, {Rieger}, {Ripken}, {Rob},
  {Rosier-Lees}, {Rowell}, {Rudak}, {Ruppel}, {Sahakian}, {Santangelo},
  {Schlickeiser}, {Sch{\"o}ck}, {Schr{\"o}der}, {Schwanke}, {Schwarzburg},
  {Schwemmer}, {Shalchi}, {Skilton}, {Sol}, {Spangler}, {Stawarz}, {Steenkamp},
  {Stegmann}, {Superina}, {Tam}, {Tavernet}, {Terrier}, {Tibolla}, {van Eldik},
  {Vasileiadis}, {Venter}, {Vialle}, {Vincent}, {Vink}, {Vivier}, {V{\"o}lk},
  {Volpe}, {Wagner}, {Ward}, {Zdziarski}, \& {Zech}}]{aharonian2009}
{Aharonian}, F., {Akhperjanian}, A.~G., {de Almeida}, U.~B., {et~al.} 2009,
  \apj, 692, 1500

\bibitem[{{Axford}(1981)}]{axford1981}
{Axford}, W.~I. 1981, in International Cosmic Ray Conference, Vol.~12,
  International Cosmic Ray Conference, 155--203

\bibitem[{{Badenes} {et~al.}(2008){Badenes}, {Hughes}, {Cassam-Chena{\"\i}}, \&
  {Bravo}}]{badenes2008}
{Badenes}, C., {Hughes}, J.~P., {Cassam-Chena{\"\i}}, G., \& {Bravo}, E. 2008,
  \apj, 680, 1149

\bibitem[{{Blandford} \& {Eichler}(1987)}]{blanford1987}
{Blandford}, R., \& {Eichler}, D. 1987, \physrep, 154, 1

\bibitem[{{Cargill} \& {Papadopoulos}(1988)}]{cargill1988}
{Cargill}, P.~J., \& {Papadopoulos}, K. 1988, \apjl, 329, L29

\bibitem[{{Chevalier} {et~al.}(1980){Chevalier}, {Kirshner}, \&
  {Raymond}}]{chevalier1980}
{Chevalier}, R.~A., {Kirshner}, R.~P., \& {Raymond}, J.~C. 1980, \apj, 235, 186

\bibitem[{{Clementini} {et~al.}(2003){Clementini}, {Gratton}, {Bragaglia},
  {Carretta}, {Di Fabrizio}, \& {Maio}}]{clementini2003}
{Clementini}, G., {Gratton}, R., {Bragaglia}, A., {et~al.} 2003, \aj, 125, 1309

\bibitem[{{Crawford} {et~al.}(2010){Crawford}, {Still}, {Schellart}, {Balona},
  {Buckley}, {Dugmore}, {Gulbis}, {Kniazev}, {Kotze}, {Loaring}, {Nordsieck},
  {Pickering}, {Potter}, {Romero Colmenero}, {Vaisanen}, {Williams}, \&
  {Zietsman}}]{crawford2010}
{Crawford}, S.~M., {Still}, M., {Schellart}, P., {et~al.} 2010, in \procspie,
  Vol. 7737, Observatory Operations: Strategies, Processes, and Systems III,
  773725

\bibitem[{{Eidelman} {et~al.}(2004){Eidelman}, {Hayes}, {Olive},
  {Aguilar-Benitez}, {Amsler}, \& {Asner}}]{pdg2004}
{Eidelman}, S., {Hayes}, K., {Olive}, K., {et~al.} 2004, {Physics Letters B},
  592

\bibitem[{{Eriksen} {et~al.}(2011){Eriksen}, {Hughes}, {Badenes}, {Fesen},
  {Ghavamian}, {Moffett}, {Plucinksy}, {Rakowski}, {Reynoso}, \&
  {Slane}}]{eriksen2011}
{Eriksen}, K.~A., {Hughes}, J.~P., {Badenes}, C., {et~al.} 2011, \apjl, 728,
  L28

\bibitem[{{Ghavamian} {et~al.}(2007){Ghavamian}, {Laming}, \&
  {Rakowski}}]{ghavamian2007b}
{Ghavamian}, P., {Laming}, J.~M., \& {Rakowski}, C.~E. 2007, \apjl, 654, L69

\bibitem[{{Giordano} {et~al.}(2012){Giordano}, {Naumann-Godo}, {Ballet},
  {Bechtol}, {Funk}, {Lande}, {Mazziotta}, {Rain{\`o}}, {Tanaka}, {Tibolla}, \&
  {Uchiyama}}]{giordano2012}
{Giordano}, F., {Naumann-Godo}, M., {Ballet}, J., {et~al.} 2012, \apjl, 744, L2

\bibitem[{{Hamuy} {et~al.}(1992){Hamuy}, {Walker}, {Suntzeff}, {Gigoux},
  {Heathcote}, \& {Phillips}}]{hamuy1992}
{Hamuy}, M., {Walker}, A.~R., {Suntzeff}, N.~B., {et~al.} 1992, \pasp, 104, 533

\bibitem[{{Helder} {et~al.}(2010){Helder}, {Kosenko}, \& {Vink}}]{helder2010}
{Helder}, E.~A., {Kosenko}, D., \& {Vink}, J. 2010, \apjl, 719, L140

\bibitem[{{Helder} {et~al.}(2011){Helder}, {Kosenko}, \& {Vink}}]{helder2011}
---. 2011, \apjl, 737, L46

\bibitem[{{Hovey} {et~al.}(2015){Hovey}, {Hughes}, \& {Eriksen}}]{hovey2015}
{Hovey}, L., {Hughes}, J.~P., \& {Eriksen}, K. 2015, \apj, 809, 119

\bibitem[{{Hughes} {et~al.}(1995){Hughes}, {Hayashi}, {Helfand}, {Hwang},
  {Itoh}, {Kirshner}, {Koyama}, {Markert}, {Tsunemi}, \& {Woo}}]{hughes1995}
{Hughes}, J.~P., {Hayashi}, I., {Helfand}, D., {et~al.} 1995, \apjl, 444, L81

\bibitem[{{Jaschek} \& {Jaschek}(1963)}]{jaschek1963}
{Jaschek}, M., \& {Jaschek}, C. 1963, \pasp, 75, 365

\bibitem[{{Lagage} \& {Cesarsky}(1983)}]{lagage1983}
{Lagage}, P.~O., \& {Cesarsky}, C.~J. 1983, \aap, 125, 249

\bibitem[{{Long} {et~al.}(1981){Long}, {Helfand}, \& {Grabelsky}}]{long1981}
{Long}, K.~S., {Helfand}, D.~J., \& {Grabelsky}, D.~A. 1981, \apj, 248, 925

\bibitem[{{Lopez} {et~al.}(2009){Lopez}, {Ramirez-Ruiz}, {Badenes},
  {Huppenkothen}, {Jeltema}, \& {Pooley}}]{lopez2009}
{Lopez}, L.~A., {Ramirez-Ruiz}, E., {Badenes}, C., {et~al.} 2009, \apjl, 706,
  L106

\bibitem[{{Massey} {et~al.}(1988){Massey}, {Strobel}, {Barnes}, \&
  {Anderson}}]{massey1988}
{Massey}, P., {Strobel}, K., {Barnes}, J.~V., \& {Anderson}, E. 1988, \apj,
  328, 315

\bibitem[{{McConnachie}(2012)}]{mcconnachie2012}
{McConnachie}, A.~W. 2012, \aj, 144, 4

\bibitem[{{Morlino} {et~al.}(2013){Morlino}, {Blasi}, {Bandiera}, \&
  {Amato}}]{morlino2013c}
{Morlino}, G., {Blasi}, P., {Bandiera}, R., \& {Amato}, E. 2013, \aap, 558, A25

\bibitem[{{O'C Drury} {et~al.}(1996){O'C Drury}, {Duffy}, \&
  {Kirk}}]{drury1996}
{O'C Drury}, L., {Duffy}, P., \& {Kirk}, J.~G. 1996, \aap, 309, 1002

\bibitem[{{Peters} {et~al.}(2013){Peters}, {Lopez}, {Ramirez-Ruiz}, {Stassun},
  \& {Figueroa-Feliciano}}]{peters2013}
{Peters}, C.~L., {Lopez}, L.~A., {Ramirez-Ruiz}, E., {Stassun}, K.~G., \&
  {Figueroa-Feliciano}, E. 2013, \apjl, 771, L38

\bibitem[{{Rest} {et~al.}(2005){Rest}, {Stubbs}, {Becker}, {Miknaitis},
  {Miceli}, {Covarrubias}, {Hawley}, {Smith}, {Suntzeff}, {Olsen}, {Prieto},
  {Hiriart}, {Welch}, {Cook}, {Nikolaev}, {Huber}, {Prochtor}, {Clocchiatti},
  {Minniti}, {Garg}, {Challis}, {Keller}, \& {Schmidt}}]{rest2005lmc}
{Rest}, A., {Stubbs}, C., {Becker}, A.~C., {et~al.} 2005, \apj, 634, 1103

\bibitem[{{Rest} {et~al.}(2008){Rest}, {Welch}, {Suntzeff}, {Oaster},
  {Lanning}, {Olsen}, {Smith}, {Becker}, {Bergmann}, {Challis}, {Clocchiatti},
  {Cook}, {Damke}, {Garg}, {Huber}, {Matheson}, {Minniti}, {Prieto}, \&
  {Wood-Vasey}}]{rest2008b}
{Rest}, A., {Welch}, D.~L., {Suntzeff}, N.~B., {et~al.} 2008, \apjl, 681, L81

\bibitem[{{Reynolds}(2008)}]{reynolds2008}
{Reynolds}, S.~P. 2008, \araa, 46, 89

\bibitem[{{Slane} {et~al.}(2014){Slane}, {Lee}, {Ellison}, {Patnaude},
  {Hughes}, {Eriksen}, {Castro}, \& {Nagataki}}]{slane2014}
{Slane}, P., {Lee}, S.-H., {Ellison}, D.~C., {et~al.} 2014, \apj, 783, 33

\bibitem[{{Slane} {et~al.}(2015){Slane}, {Lee}, {Ellison}, {Patnaude},
  {Hughes}, {Eriksen}, {Castro}, \& {Nagataki}}]{slane2015}
---. 2015, \apj, 799, 238

\bibitem[{{Smith} {et~al.}(1991){Smith}, {Kirshner}, {Blair}, \&
  {Winkler}}]{smith1991}
{Smith}, R.~C., {Kirshner}, R.~P., {Blair}, W.~P., \& {Winkler}, P.~F. 1991,
  \apj, 375, 652

\bibitem[{{Smith} {et~al.}(1994){Smith}, {Raymond}, \& {Laming}}]{smith1994}
{Smith}, R.~C., {Raymond}, J.~C., \& {Laming}, J.~M. 1994, \apj, 420, 286

\bibitem[{{Stone}(1996)}]{stone1996}
{Stone}, R.~P.~S. 1996, \apjs, 107, 423

\bibitem[{{Tuohy} {et~al.}(1982){Tuohy}, {Dopita}, {Mathewson}, {Long}, \&
  {Helfand}}]{tuohy1982}
{Tuohy}, I.~R., {Dopita}, M.~A., {Mathewson}, D.~S., {Long}, K.~S., \&
  {Helfand}, D.~J. 1982, \apj, 261, 473

\bibitem[{{van Adelsberg} {et~al.}(2008){van Adelsberg}, {Heng}, {McCray}, \&
  {Raymond}}]{vanadelsberg2008}
{van Adelsberg}, M., {Heng}, K., {McCray}, R., \& {Raymond}, J.~C. 2008, \apj,
  689, 1089

\bibitem[{{van Dokkum}(2001)}]{vandokkum2001}
{van Dokkum}, P.~G. 2001, \pasp, 113, 1420

\bibitem[{{Yamaguchi} {et~al.}(2014){Yamaguchi}, {Badenes}, {Petre}, {Nakano},
  {Castro}, {Enoto}, {Hiraga}, {Hughes}, {Maeda}, {Nobukawa}, {Safi-Harb},
  {Slane}, {Smith}, \& {Uchida}}]{yamaguchi2014}
{Yamaguchi}, H., {Badenes}, C., {Petre}, R., {et~al.} 2014, \apjl, 785, L27

\end{thebibliography}
\end{document}